\documentclass[fleqn,10pt]{wlscirep}
\usepackage[utf8]{inputenc}
\usepackage[T1]{fontenc}

\usepackage{subcaption}
\usepackage{amsmath}
\usepackage{pdflscape}
\usepackage{lineno}

\title{Properties of quantum emitters in different hBN sample types particularly suited for nanophotonic integration}

\author[1]{Ambika Shorny}
\author[1]{Hardy Schauffert}
\author[2]{James C.Stewart}
\author[3]{Sajid Ali}
\author[1,5]{Stefan Walser}
\author[1]{Helmut Hörner}
\author[1]{Adarsh S. Prasad}
\author[2]{Vitaly Babenko}
\author[2]{Ye Fan}
\author[4]{Dominik Eder}
\author[3]{Kristian S. Thygesen}
\author[2]{Stephan Hofmann}
\author[4]{Bernhard C. Bayer}
\author[1,*]{Sarah M. Skoff}

\affil[1]{Atominstitut, Technische Universit\"at Wien, Stadionallee 2, Vienna, 1020, Austria}

\affil[2]{Department of Engineering, University of Cambridge, 9 JJ Thomson Avenue, Cambridge, CB3 0FA,  UK}

\affil[3]{Computational Atomic-scale Materials Design (CAMD), Department of Physics, Technical University of Denmark, Fysikvej, 307, Kongens Lyngby, DK-2800,  Denmark}

\affil[4]{Institute of Materials Chemistry, Technische Universit\"at Wien, Getreidemarkt 9/165, Vienna, 1060,  Austria}

\affil[5]{Institut für Experimentalphysik, Universität Innsbruck, Technikerstrasse 25/4, A-6020 Innsbruck, Austria}

\affil[*]{sarah.skoff@tuwien.ac.at}

\begin{abstract}
Single photon emitters in two-dimensional (2D) hexagonal boron nitride (hBN) are promising solid-state quantum emitters for photonic applications and quantum networks. Despite their favorable properties, much is still unknown about their characteristics and their atomic origin.  We focus on two different kinds of hBN samples that particularly lend themselves for integration with nanophotonic devices, multilayer nanoflakes produced by liquid phase exfoliation (LPE) and a layer-engineered sample from  hBN grown by chemical vapour deposition (CVD). We investigate their inherent defects and fit their emission properties to computationally simulated optical properties of likely carbon-related defects.  Thereby we compare and elucidate the properties in different sample types particularly suited for photonic quantum networks and narrow down the origin of emitters found in these samples. Our work is thus an important step towards harnessing the full potential of single photon emitters in hBN. 
\end{abstract}
\begin{document}

\flushbottom
\maketitle
%
%
\thispagestyle{empty}

\section{Introduction}\label{sec1}

Quantum emitters in two-dimensional (2D) materials are attracting much interest due to their remarkable properties\cite{tran_quantum_2016, wang_site-specific_2024,caldwell_photonics_2019,gottscholl_sub-nanoscale_2021,vogl_room_2017,schell_non-linear_2016}. Particularly, quantum emitters in hexagonal boron nitride (hBN) have been shown to be very stable over a wide temperature range \cite{kianinia_robust_2017,liu_ultrastable_2020,fournier_position-controlled_2021}, have bright emission into the zero-phonon line (ZPL) even at room temperature with Debye-Waller (DW) factors exceeding 0.8 \cite{tran_quantum_2016} and their transition frequency can be tuned via the Stark effect \cite{nikolay_very_2019}. Some quantum emitters in hBN may even exhibit lifetime-limited emission at room temperature \cite{hoese_mechanical_2020,dietrich_solid-state_2020}, which has not been seen in any other solid-state quantum emitter so far. In addition, the 2D geometry of the host material particularly lends itself for the integration with nanophotonic devices. For photonic quantum technologies, such as single photon sources, the bright quantum emitters in hBN around transition energies of 2.0 eV are particularly suited but so far much remains unknown about the atomic origin and properties of these single photon emitters \cite{bourrellier_bright_2016}. To fully exploit all of the advantages of solid-state emitters as single photon sources or constituents of photonic quantum networks and implement scalable devices, near-field coupling of the emission to waveguides or microcavities is key. 

Here, we aim to shed more light on these quantum emitters by comparing the experimental emission properties of sample types that are particularly suited for integration with nanophotonics. 
On one hand side we have chosen layer-engineered hBN films grown by chemical vapour deposition (CVD) and on the other hand commercially available small hBN flakes produced by liquid-phase exfoliation (LPE). 

The CVD grown hBN film sample is designed to work for planar waveguide chips and photonic circuits. Here the focus lies on obtaining high quality extremely thin hBN films,  more details on the production can be found in \cite{stewart_quantum_2021}. Such a layer-engineering approach allows for more control over the position of the emitters and as the thickness of the sample is only three atomic layers, light scattering by the host crystal is negligible.

The other type of sample that was investigated are LPE hBN nanoflakes in liquid suspension that can be commercially bought and have not been post-treated. Such tiny flakes are an ideal sample for integration with free-standing nanophotonic waveguides, such as optical nanofibers \cite{skoff_optical-nanofiber-based_2018}, that are naturally integrated with an optical fiber network, as they are easily deposited on such structures.   
\begin{figure}[!ht]
	\centering
	\includegraphics[width=0.7\linewidth]{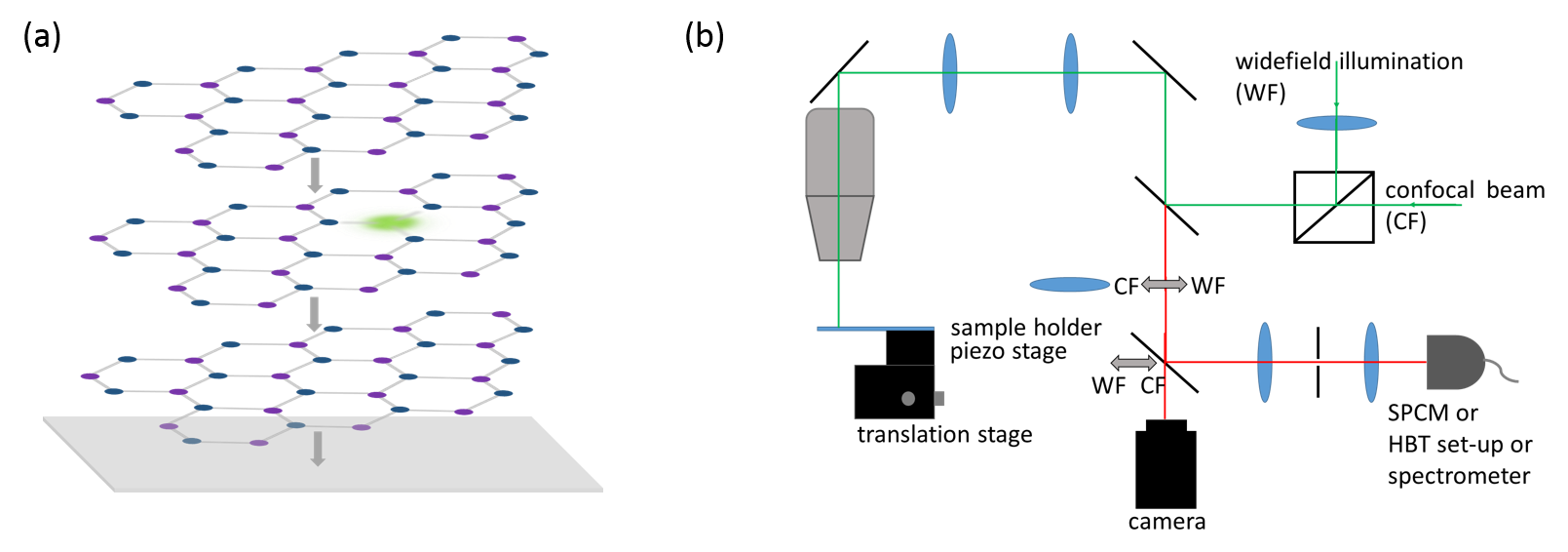}
    \caption{ (a) Sketch of the layer-engineered CVD grown hBN sample. An optically active hBN layer is sandwiched between two hBN protection layers. (b) Sketch of the confocal microscope set-up for investigating qantum emitters in hBN.} 
	\label{fig:optical_setup_cvd}
\end{figure}
 For both hBN samples we study the properties of their most abundant single photon emitters and we verify that their photoluminescence (PL) signatures correspond to the most abundant type found in the respective substrates also in previous literature \cite{chejanovsky_structural_2016,tran_quantum_2016,preus_assembly_2021,nguyen_nanoassembly_2018,kim_design_2018,schell_quantum_2018,mendelson_engineering_2019}. The investigated defects  also occur inherently in these types of samples even without any post-processing and thus we investigate their characteristics and narrow down their possible origin by comparing computationally simulated PL properties of the most likely defect types with our experimental data from the two samples.

\section{Results and Discussion}\label{sec2}

\subsection{Layer-engineered hBN}

For integration with chip-based photonic structures, a thin, flat material is advantageous. For this purpose a stack of three CVD hBN layers is engineered \cite{stewart_quantum_2021}, where only the central layer hosts single photon emitters and the two outer layers function as a protection against bleaching of the emitters (Fig. \ref{fig:optical_setup_cvd}a). This is achieved by transferring three CVD grown monolayers, that have been cleaned of polymer residue by annealing in air at 450° for about 20 min, on top of each other. Only the central layer hosts quantum emitters, which have been re-introduced by Ar annealing at about 850°C for 30 min.  It has to be noted that these emitters that were re-introduced via Ar annealing were the same in nature as those that were inherently found beforehand. More details on the sample preparation can be found in the Methods section and in Stewart et al. \cite{stewart_quantum_2021}.  

\begin{figure}[hb]
	\centering
	\includegraphics[width=0.5\linewidth]{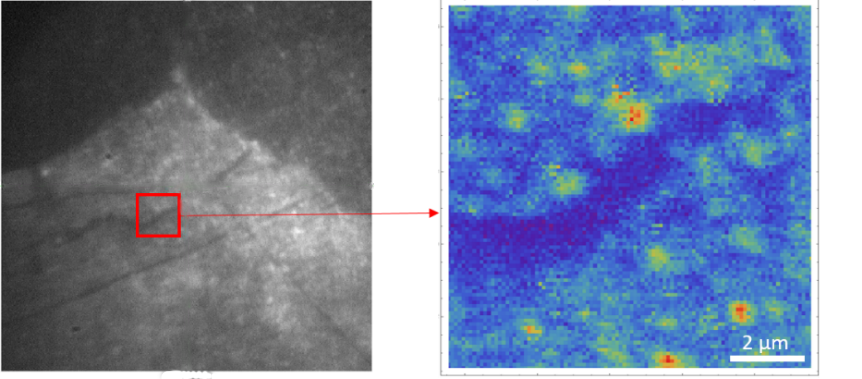}
    \caption{Widefield image of a trilayer sample of hBN (left) and corresponding confocal image (right) of the marked area} 
	\label{fig:wfandcfnew}
\end{figure}

We study this trilayer sample using a homebuilt confocal microscope (Fig. \ref{fig:optical_setup_cvd} with a diode laser at 532 nm. The microscope can either function in confocal or widefield mode, an example of which can be seen in Fig. \ref{fig:wfandcfnew}. 

\begin{figure}[h!]
\centering
\includegraphics[width=0.9\linewidth]{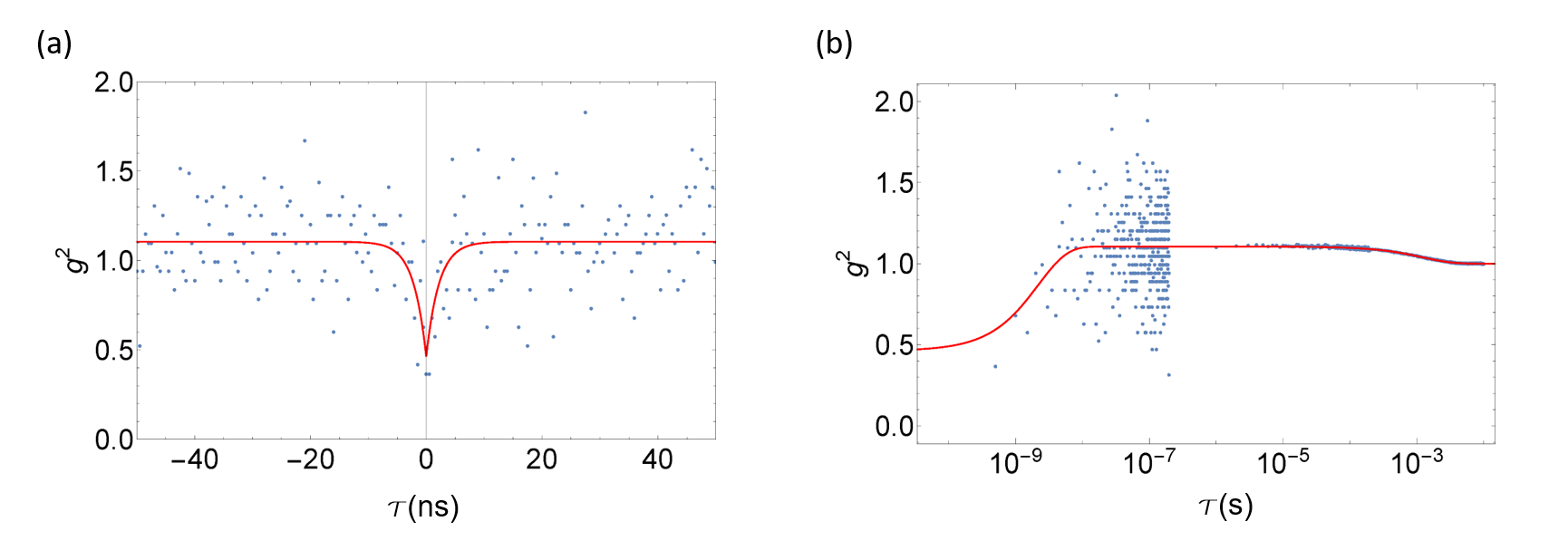}
\caption{(a) Second order intensity correlation measurements and corresponding fit (red line) showing the typical antibunching dip of a single photon emitter. (b) For longer correlation times, photon bunching due to at least one metastable state can be observed.}
\label{fig:cvdg2shortlong}
\end{figure}

In both cases we collect the fluorescence of the sample and hence regions with quantum emitters appear bright. The collected light is then sent to a Hanbury-Brown-Twiss (HBT) set-up to check for single photon emission or to a spectrometer to evaluate the PL profile. Figure \ref{fig:cvdg2shortlong} shows a second-order intensity correlation measurement of an emitter in layer-engineered hBN showing antibunching for a time delay of $\tau =0$ and thus proving that it is indeed a single photon emitter. Investigating these correlations over longer times shows that the emitter we are looking at has at least one metastable energy level. More information on these measurements and detailed information on the fit is given in the Supplemental Information.   

\subsection{hBN quantum emitters in liquid phase exfoliated nanoflakes}

\begin{figure}[b!]
	\centering
	\includegraphics[width=0.5\linewidth]{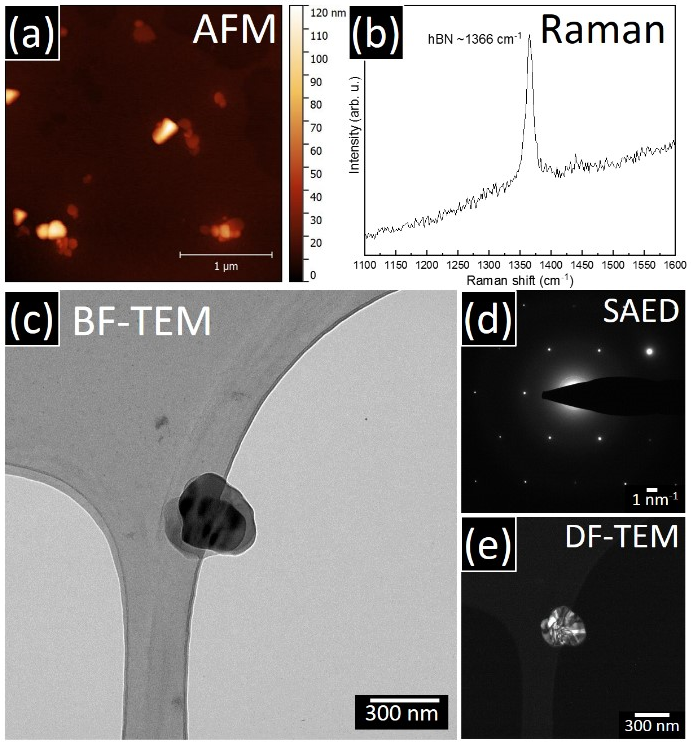}
    \caption{(a) AFM micrograph of LPE hBN platelets drop cast from suspension onto SiO$_2$ covered Si wafer. (b) Raman spectrum corresponding to (a). (c) BF-TEM, (d) SAED and (e) DF-TEM images of a single hBN flake deposited onto a lacey carbon TEM grid.} 
	\label{fig:characterization}
\end{figure}
The other type of hBN sample investigated for typical quantum emitters that emit around 2 eV can be purchased commercially and is useful for dropcasting on photonic components or interfacing with free-standing waveguides as we demonstrate below. 

The LPE nanoflakes purchased from Graphene Supermarket are suspended in a water:ethanol mixture. To estimate their lateral size and thickness distributions and confirm their structure Atomic force microscopy (AFM),  Raman spectroscopy and transmission electron microscopy (TEM) in bright-field (BF) and dark-field (DF) mode and selected area electron diffraction (SAED) are used. For that purpose a droplet of the solution is dropcast onto a SiO$_2$ covered Si wafer or lacy carbon TEM grid.

\begin{figure}
\centering
\begin{subfigure}[c]{0.4\textwidth}
\includegraphics[width=\linewidth]{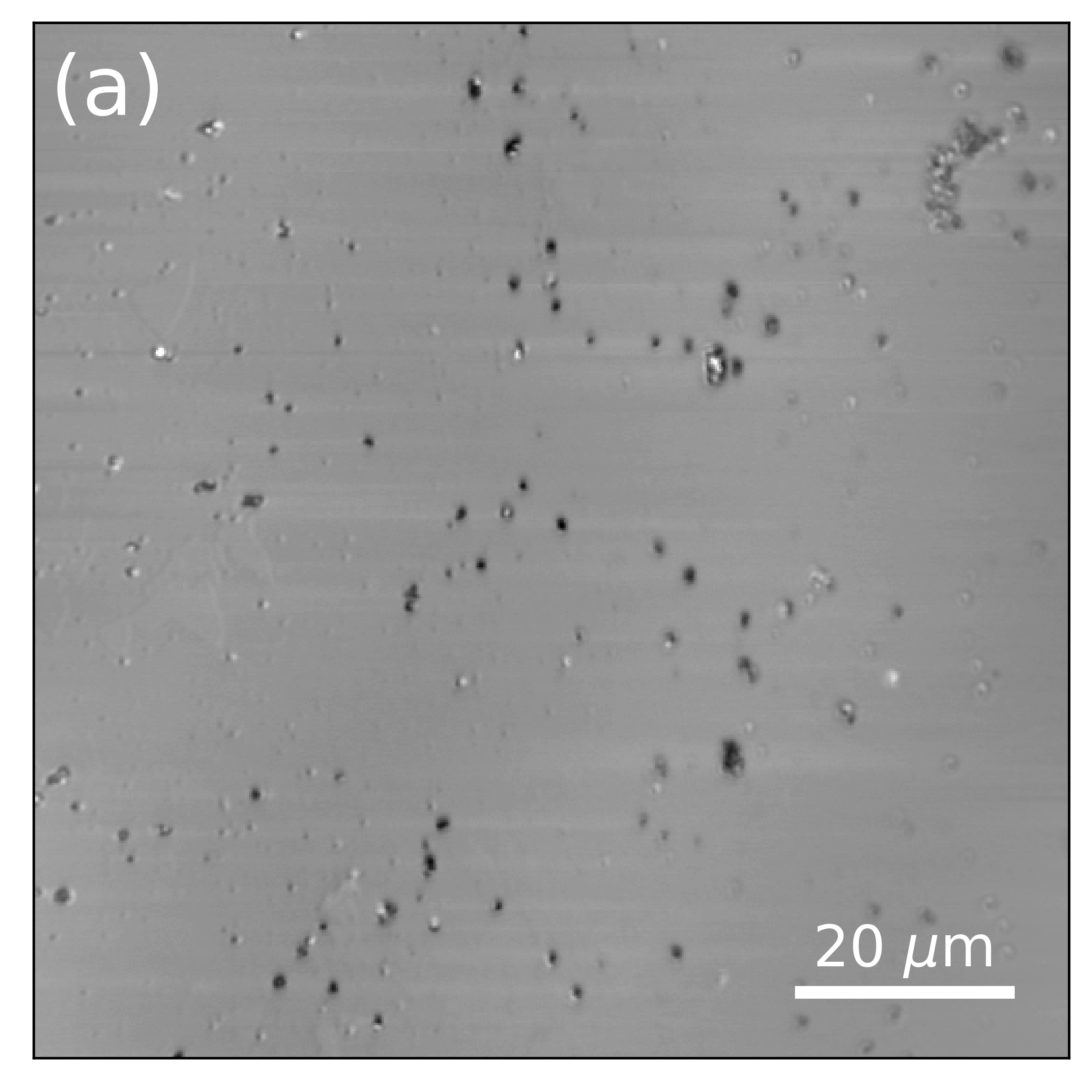}
\end{subfigure}
\begin{subfigure}[c]{0.4\textwidth}
\includegraphics[width=\linewidth]{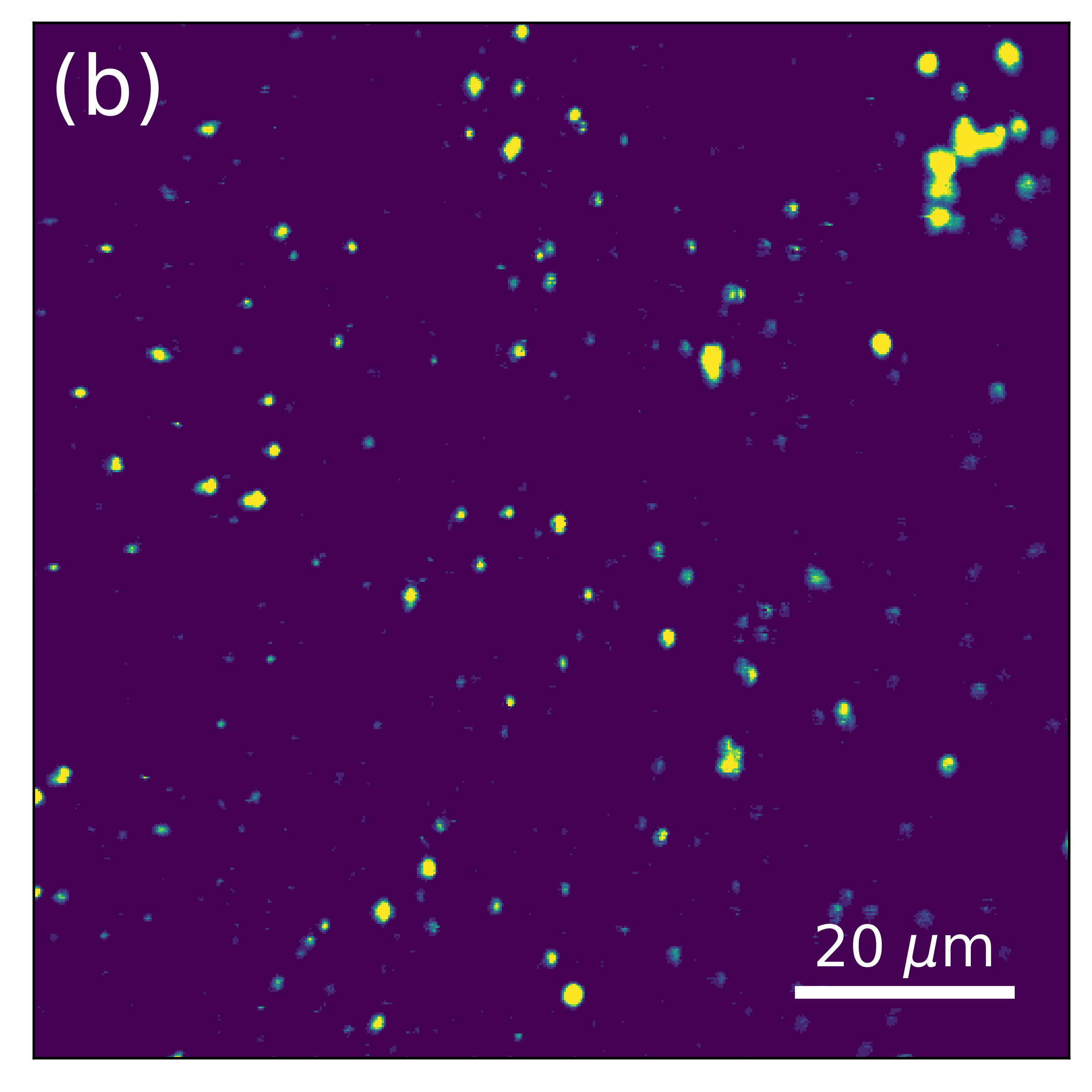}
\end{subfigure}
\caption{(a) Confocal reflection microscopy image of LPE flakes on glass and (b) the corresponding fluorescence image }
\label{fig:LPEconfocal}
\end{figure}

 From AFM height profiles (Fig. \ref{fig:characterization}a) of individual flakes, we estimate a mean of 160 nm  from the lateral platelet size and a mean of 28 nm for their height with a thickness:length ratio of the flakes of 0.16 confirming their platelet-type morphology. Considering that a hBN monolayer has a height of 0.34 nm \cite{caneva_nucleation_2015} the observed flakes are few- to multi-layered hBN. The Raman spectrum in Fig. \ref{fig:characterization}b confirms that the platelets are hBN via its characteristic Raman signal at $\sim$1366 cm$^{-1}$ \cite{caneva_nucleation_2015}. The BF-TEM in Fig. \ref{fig:characterization}c shows a single hBN platelet deposited onto a lacey carbon strand. The corresponding SAED pattern in Fig. \ref{fig:characterization}d confirms that this hBN platelet is a hBN single crystal \cite{bayer_introducing_2017}. The DF-TEM in Fig. \ref{fig:characterization}e however indicates via contrast variation that within the hBN single crystal stacking faults exist \cite{kim_stacking_2013}. Notably, both on flat SiO$_2$ covered Si wafer and on more curved lacey carbon supports all observed hBN flakes are stuck to the support in planar configuration, suggesting that a similar planar hBN orientation on an optical waveguide surface should be feasible.
The LPE hBN is also investigated via our confocal microscope, where an example image of flakes deposited on a SiO$_2$ substrate is shown in Fig. \ref{fig:LPEconfocal}. 

Fig. \ref{fig:LPEconfocal} (a) shows an image of the backscattered laser beam, depicting the LPE hBN flakes as shadows, while in (b) we use fluorescence excitation microscopy to visualize the bright emitters within the flakes. Here, no post-processing to the LPE hBN has been used, if one wishes to increase the density of stable emitters, post-processing of the sample like plasma treatment or annealing can be performed \cite{li_prolonged_2023, zeng_single-photon_2024}. Here, we wanted to investigate the emitters that are present even without this post-treatment. From Fig. \ref{fig:LPEconfocal} we can evaluate the fraction of flakes that show a fluorescence signal and find that about 53\% of flakes show fluorescence. This emission can again be analyzed for single photon emission via the HBT set-up and for its spectral properties by sending it to a spectrometer.

\begin{figure}[h!]
\centering
\begin{subfigure}[c]{0.5\textwidth}

\includegraphics[width=\linewidth]{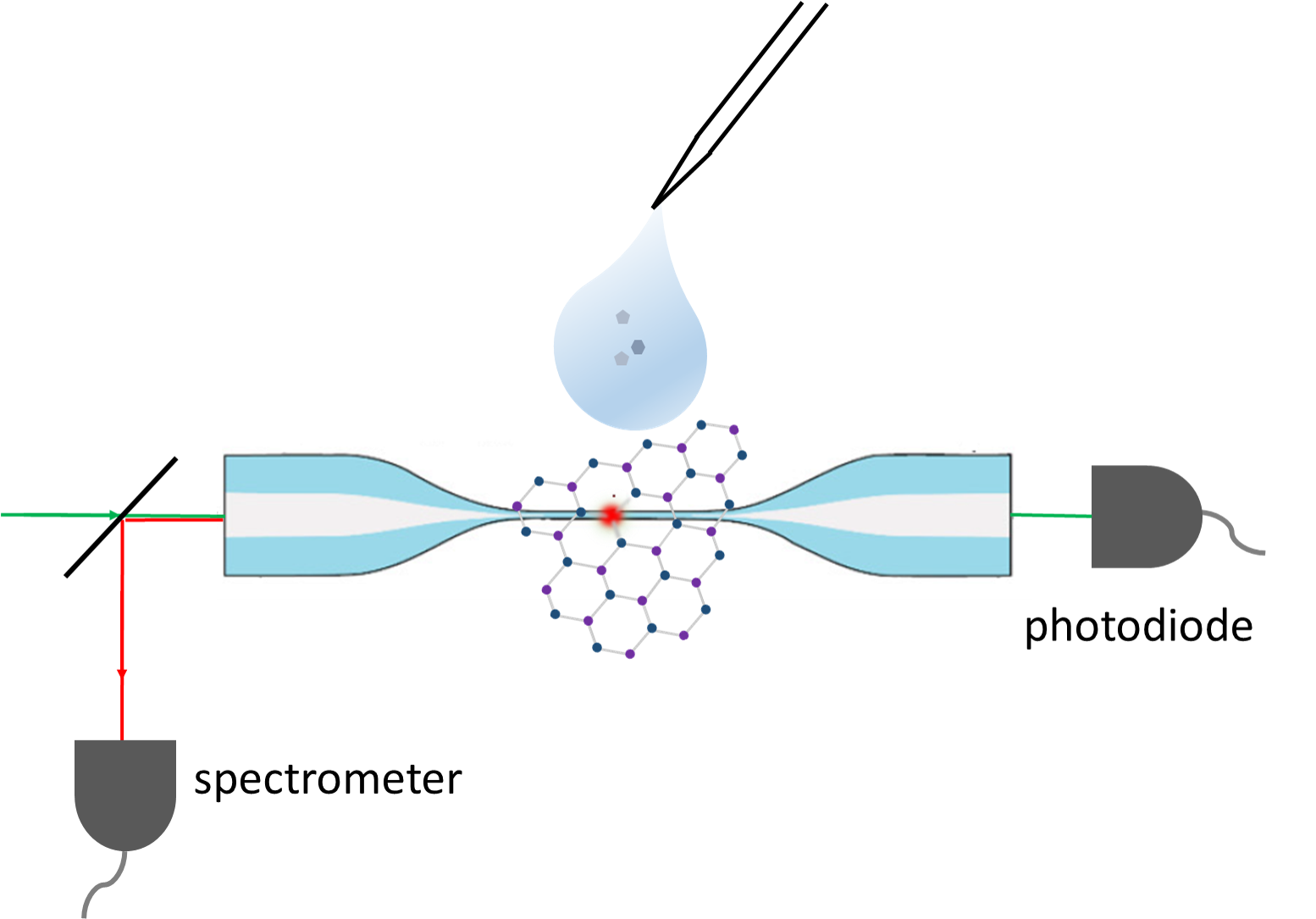}

\end{subfigure}
\begin{subfigure}[c]{0.3\textwidth}
\includegraphics[width=\linewidth]{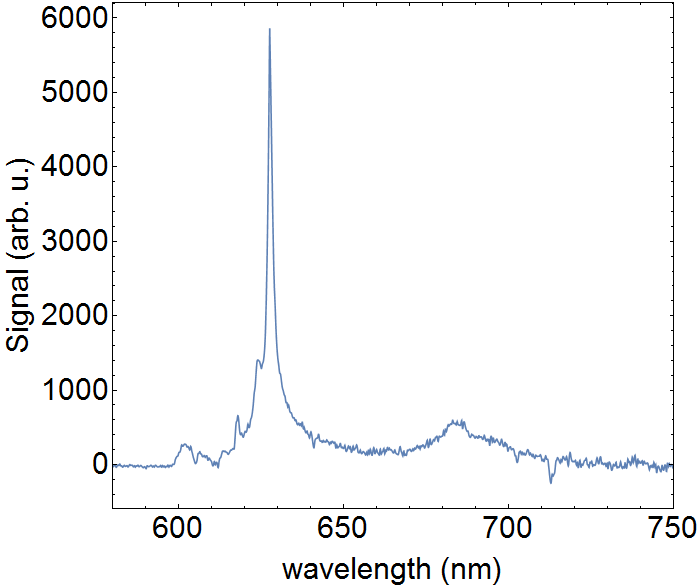}
\end{subfigure}
\caption{Sketch of the drop-touch method used (left). The fluorescence from and the transmission of the nanofiber is continuously monitored during the deposition process. On the right, the spectrum of a single quantum emitter coupled to the light field of an optical nanofiber as displayed by the spectrometer, is shown.}
\label{fig:optical_setup_fiber_spectrum}
\end{figure}

\subsubsection{LPE hBN interfaced with an optical nanofiber}
To demonstrate that even the commercial LPE sample is well suited for nanophotonic integration, we have interfaced a nanoflake with the guided light of an optical nanofiber and also performed spectroscopy via that platform (Fig. \ref{fig:optical_setup_fiber_spectrum}). Optical nanofibers offer a convenient way for near-field coupling of quantum emission. By placing solid-state emitters on the surface of these nanophotonic devices, coupling efficiencies of over 20 \% can be achieved \cite{skoff_optical-nanofiber-based_2018,shafi_efficient_2020,yalla_efficient_2012, schell_coupling_2017} and when combined with cavity structures \cite{schell_highly_2015}, Purcell factors of more than 15 are possible \cite{hutner_nanofiber-based_2020,takashima_fabrication_2019}. We employ optical nanofibers that are tapered by a custom made heat-and-pull process \cite{warken_fiber_2008}. A nanofiber waist of 320 nm ensures that for an excitation wavelength of 570 nm, only the fundamental HE11 mode is guided. Emission into a single mode is a pre-requisite for many applications of future quantum technologies and is also needed for fundamental experiments in quantum optics.
\begin{figure}[ht!]
\centering
\includegraphics[width=0.9\linewidth]{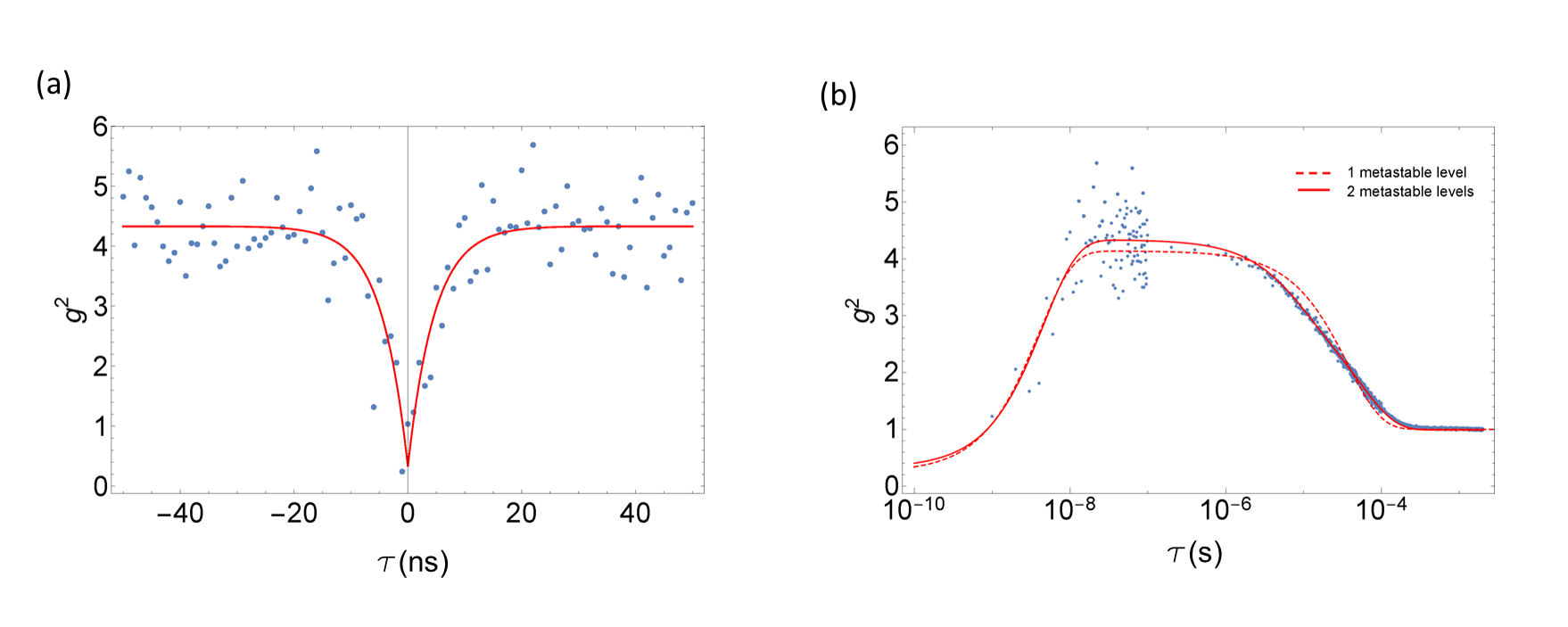}
\caption{(a) Second order intensity correlation measurements showing the typical antibunching dip of a single photon emitter and the corresponding fit to a rate equation model.(b) For longer correlation times, photon bunching due to at least two metastable states can be observed}
\label{fig:lpeg2shortlong}
\end{figure}
The hBN flakes are placed on the nanofiber by a “drop-touch” method. Hereby a drop from a syringe of hBN suspension is put into contact with the nanofiber surface using a micro-translation stage and then moved away again, until a flake adheres to the nanofiber surface. During this process, the light of a dye laser at 570 nm is coupled into the optical fiber and the backscattered fluorescence is measured using a spectrometer (Fig. \ref{fig:optical_setup_fiber_spectrum} a). Reflected laser light is filtered out by a longpass filter for 610 nm. If an hBN flake containing quantum emitters has been deposited on the nanofiber in this process, a fluorescence from the quantum emitters can be seen in the fluorescence spectrum. If not, the process can be repeated. This process of deposition of nanoparticles on a nanofiber is simple and scaleable. We have also demonstrated that single emitters can in this way be coupled to the guided light field of the waveguide as we have performed g$^2$ intensity correlation measurements proving single photon emission and corresponding spectral measurements via the nanofiber platform as can be seen in Fig. \ref{fig:optical_setup_fiber_spectrum} b and Fig. \ref{fig:lpeg2shortlong} a.  Looking at longer correlation times, the fit to a four level rate equation model in Fig. \ref{fig:lpeg2shortlong} b reveals that the probed emitter exhibits at least two metastable levels. More information including the fit results of the correlation measurements is given in the Supplemental Section.

\subsection{Comparison of experimental and theoretical spectra}

To gain more information on the types of defects responsible for the single photon emission in the two chosen hBN sample types, we investigate their spectral emission characteristics in detail. 
When exciting the single photon emitters with energies higher than the resonance electronic transition, the  PL spectrum consists of both the zero phonon line (ZPL) and the phonon sideband (PSB). Such a spectrum therefore yields important information on the electronic and phononic transitions of the defect and hence its potential energy surface. The expected transitions can also be theoretically simulated and thus we compare our experimental results to those of simulated defects in hBN. 
\begin{figure}[ht]
	\centering
	\includegraphics[width=0.8\linewidth]{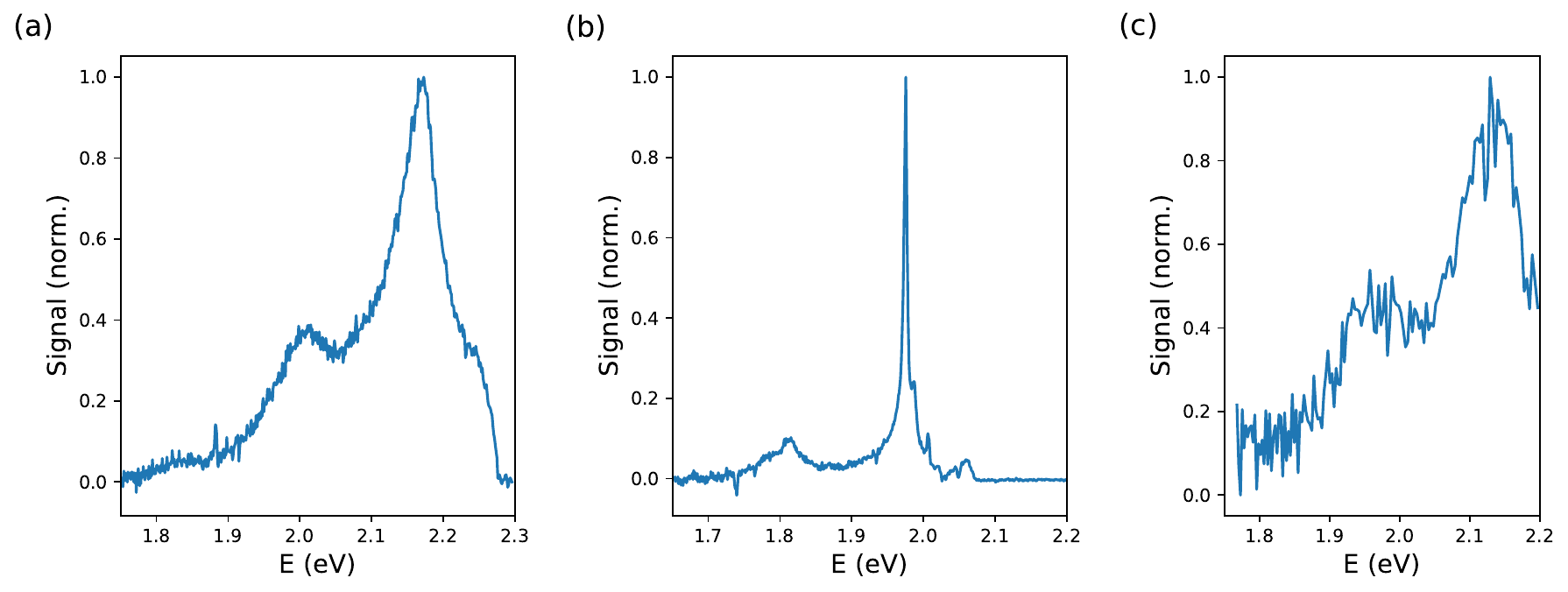}
    \caption{PL spectra of defects found in  (a) CVD grown, layer-engineered hBN and (b,c) LPE hBN samples} 
	\label{fig:CVDLPEspectra}
\end{figure}
In Fig. \ref{fig:CVDLPEspectra}, typical spectra of inherent emitters in CVD hBN and LPE hBN are shown. Whereas CVD hBN inherently seems to host one type of emitter (Fig. \ref{fig:CVDLPEspectra} a), LPE hBN hosts emitters with broad and narrow emission (Fig. \ref{fig:CVDLPEspectra} b and c)
 Comparing the two emitter types in LPE hBN shows that the narrow emitters are much more stable and robust against photobleaching. By fitting both the ZPL and PSB to simple Lorentzians, we can make a first estimate on the spectral width, transition energy and energy of the phonon modes (Tab. \ref{tab:fitspectralorentzian}). 
A spectral width of the ZPL of around 70 meV for single emitters in CVD hBN at room temperature that are predominantly found at a transition energy of 2.15 eV is also what has been previously reported in literature for standard CVD growth conditions \cite{chejanovsky_structural_2016,abidi_selective_2019,comtet_wide-field_2019,glushkov_direct_2021}. LPE hBN has been shown to host single photon emitters with narrower emission and we find an average linewdith of 16 meV at predominantly lower transition energy of 1.97 eV. This is in agreement with previous values obtained  for stable emitters in LPE hBN in literature using the same fitting procedure \cite{kim_integrated_2019,schell_quantum_2018,nguyen_nanoassembly_2018}. We  have also found this narrow emission at 2.14 eV and 1.68 eV as indicated in Tab.\ref{tab:fitspectralorentzian}. Broad emission in LPE hBN is also observed but is much less stable than that found in CVD hBN. In fact the bleaching occurs within a few seconds and thus the single emitter character of this fluorescence could not be verified for this type of emission.  A reason why these emitters are not commonly seen in other literature apart from a recent paper \cite{li_prolonged_2023} is that they vanish when annealing the samples under an argon atmosphere, which is a common process performed before most experiments are conducted. Kretzschmar et al.\cite{kretzschmar_quantitative_2024} compare  emitters in different commercial LPE hBN flakes and the ratio of flakes with emitters to the total number of flakes is evaluated. For the same supplier as was used in our experiment, they obtain a fraction of 10\%, whereas we consistently see > 50\% of the flakes showing fluorescence (Fig.\ref{fig:LPEconfocal}) . The reason for this is likely that they have bleached all the broad emission when heating the samples for 15 minutes to 870° in an Ar atmosphere and are therefore left with the narrow, stable emitters. This is also underpinned by the results in \cite{li_prolonged_2023}, where annealing in an Argon atmosphere gets rid of the broad emitters and leaves the stable narrow emitters unaffected.

\begin{table}[b!]
\centering
\begin{tabular}{{|p{2.5cm}||p{3.5cm}|p{3.5cm}|p{3.5cm}|}}
\hline
&{\centering\text{$x_{0,\text{ZPL}}$ [eV]}}&{\text{\centering$\gamma$ (FWHM) [eV]}}&{\centering$x_{0,\text{ZPL}}-x_{0,\text{PSB}}$ [eV]}\\
\hline
\hline

CVD & 2.166$\pm$0.024 &0.073$\pm$0.013 &0.153$\pm$0.013 \\
\hline
LPE narrow &  $1.968\pm0.012$ &$0.016\pm0.007$ & 0.156$\pm$0.022 \\
\hline
LPE narrow&  $2.142\pm 0.001$ (fit) &$0.011\pm 0.001$ (fit)& $0.163\pm0.001$ (fit) \\
\hline
LPE narrow&  $1.677\pm0.001$ (fit)&$0.017\pm0.001$ (fit)& * \\
\hline
LPE broad &  $2.139\pm 0.006$ & $0.133\pm0.007$ & $0.178 \pm 0.001$ \\

\hline
\end{tabular}
\caption{Results of fitting the PL of different samples to simple Lorentzians. The narrow emission for the LPE samples was found at three distinct transition energies whereas the emission in CVD hBN and for the broad emission in LPE hBN was centered around 2.15 eV .  The results and error correspond to the mean and standard deviation of the emission found in the various samples unless indicated by (fit), which means that the fit results of an individual spectrum are displayed. * Here the PSB wasn't recorded by the spectrometer as the ZPL was so far red-shifted}
\label{tab:fitspectralorentzian}
\end{table}

\begin{figure}[ht]
	\centering
	\includegraphics[width=0.9\linewidth]{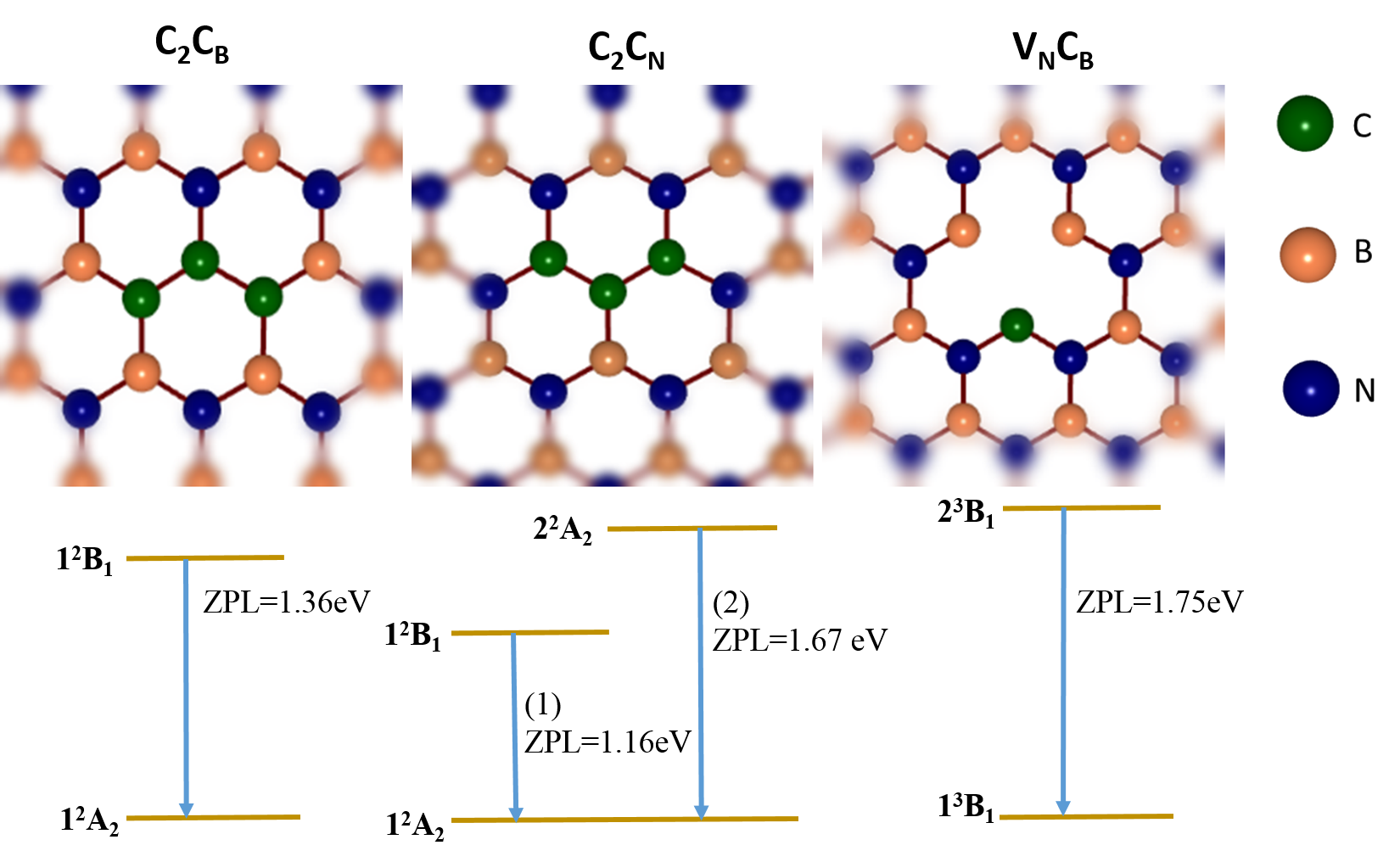}
    \caption{Graphical representation of the defects  $\text{V}_\text{N}\text{C}_\text{B}$, $\text{C}_\text{2}\text{C}_\text{N}$ and $\text{C}_\text{2}\text{C}_\text{B}$ with $\text{C}_\text{2v}$ symmetry embedded in a periodic hBN sheet. Beneath each defect, the relevant calculated transitions that are candidates for single photon emission in the visible range, are shown.} 
	\label{fig:figdefects}
\end{figure}

Theoretically, a number of possible candidates for single photon emitters in hBN has been suggested \cite{stewart_quantum_2021, sajid_defect_2018} from simple vacancies \cite{fischer_controlled_2021} to carbon-based defects \cite{mendelson_identifying_2021}, to dangling bonds \cite{turiansky_boron_2021}.
There are conditions that these defects have to meet for them to be likely to be the origin of single photon emitters in hBN in the visible range.
\begin{figure}[htb!]
	\centering
	\includegraphics[width=1\linewidth]{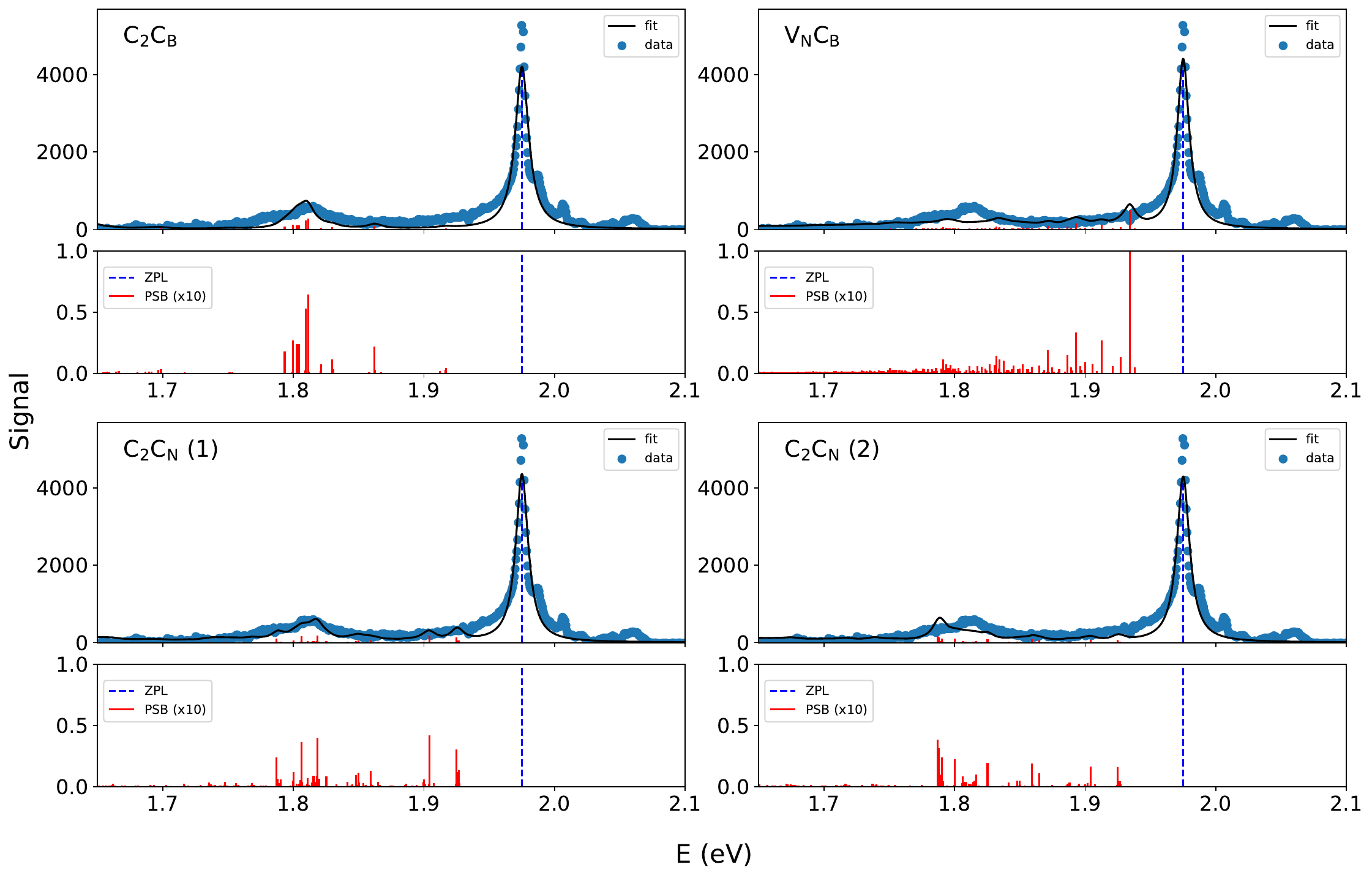}
    \caption{PL spectra of defects from LPE hBN samples fit (solid black line) to the different simulated emitters with the ZPL position (blue, dashed line) and the phonon modes (red, solid lines) shown. For better visibility, beneath each spectrum, the transitions for each defect type are displayed normalised to the strength of the ZPL and with the PSB strength multiplied by 10.} 
	\label{fig:expfit_theory_spectrum_lpe}
\end{figure}
It is required that at least two states of the defect are found within the bandgap of hBN that are neither both fully occupied nor both completely empty and that the calculated PL spectrum  features a  sharp and strong ZPL with a phonon sideband peak red-shifted by about 150-180 meV from the ZPL as observed experimentally. Density functional theory (DFT) calculations are usually not very exact concerning the absolute transition energy and thus the results of the absolute ZPL energy are not trustworthy enough to pick the correct origin of the emitter. However, the spacing between different transitions is more reliable and can be used to identify possible candidates.
\begin{figure}[ht!]
	\centering
	\includegraphics[width=1\linewidth]{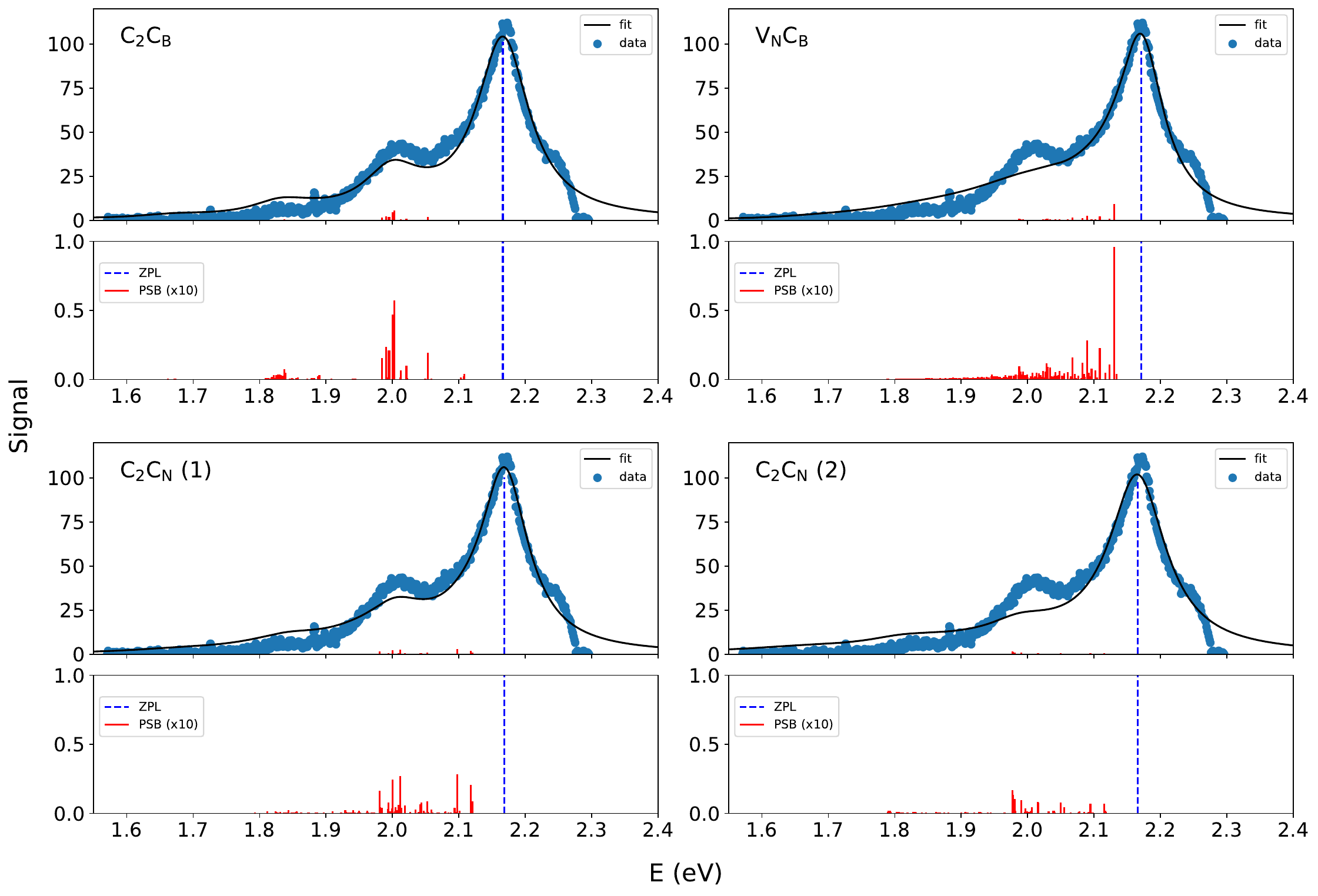}
    \caption{PL spectra of defects from CVD hBN samples fit (solid black line) to the different simulated emitters with the ZPL position (blue, dashed line) and the phonon modes (red, solid lines) shown. For better visibility, beneath each spectrum, the transitions for each defect type are displayed normalised to the strength of the ZPL and with the PSB strength multiplied by 10.} 
	\label{fig:expfit_theory_spectrum_cvd}
\end{figure}

Considering vacancies where only one atom is missing from the lattice, the $\text{V}_{\text{B}}$ defect has been identified as being likely responsible for a very broad emission at 850 nm \cite{gottscholl_initialization_2020,gottscholl_sub-nanoscale_2021,toledo_electron_2018} and hence we exclude it as a likely candidate for the ones found in the visible range that feature a narrow distinct ZPL. Although not found experimentally, calculations for the $\text{V}_{\text{N}}\text{N}_{\text{B}}$ defect have shown that this defect is also likely to feature a very broad and dominant PSB \cite{jara_first-principles_2021} and thus we do not consider this defect further here. As we are interested in emitters that are inherently found in the material without complex post-processing we also have to consider which defects are favourable concerning their formation energy. Recently it has been suggested that carbon plays a crucial role in emitter formation \cite{mendelson_identifying_2021} and it has been observed that a deliberate carbon atmosphere during the growth process indeed increases the density of emitters in CVD-grown hBN \cite{liu_quantum_2021}. Further, it has been shown that defects created by carbon ion implantation show emission around 580 nm \cite{venturi_selective_2024} also suggesting that a carbon-related defect may be the origin. Although it has been calculated \cite{jara_first-principles_2021} that the formation energy of a carbon atom substituting a boron or nitrogen atom in pristine hBN is too high to occur for thermal mechanisms, the formation energy for a carbon impurity taking the place of a pre-exisiting vacancy is very favorable. While the carbon monomers $\text{C}_\text{N}$ and $\text{C}_\text{B}$ have been shown to be unsuitable as candidates for single photon emitters, the carbon trimers $\text{C}_\text{2}\text{C}_\text{B}$ and $\text{C}_\text{2}\text{C}_\text{N}$ (where $\text{C}_\text{2}$ stands for $\text{C}_\text{N}\text{C}_\text{B}$) are likely candidates. The formation energy needed for these impurities to fill a vacancy are found to be -9.12 eV and -5.6 eV \cite{jara_first-principles_2021}, for the $\text{C}_\text{2}\text{C}_\text{B}$ and $\text{C}_\text{2}\text{C}_\text{N}$, respectively and thus this process is very favorable in both cases.

Therefore, we focus on three different possible defects, the  $\text{C}_{2}\text{C}_{\text{B}}$, the $\text{C}_2\text{C}_{\text{N}}$ and the $\text{V}_{\text{N}}\text{C}_{\text{B}}$ defect. The schematic atomic structure of the three defects in $\text{C}_{\text{2v}}$ symmetry is shown in Fig. \ref{fig:figdefects}.
For our calculations of the PL spectra of these three defects, we consider ZPL transitions that have energies between 1-3 eV as uncertainties on the order of 0.5-1 eV are not uncommon for absolute results of the transition energies from DFT simulations.  
The relevant transitions are displayed in Fig. \ref{fig:figdefects}. For $\text{C}_2\text{C}_{\text{N}}$ two possible transitions exist that meet the above criteria and these are labelled (1) and (2) throughout the manuscript corresponding to the $1 ^{2}B_1 \rightarrow 1 ^{2}A_2$ and the $2 ^{2}A_2 \rightarrow 1 ^{2}A_2$ transition, respectively.
The planar configuration of the dipole for ground and excited state is the most stable one for all cases except for the excited state of $\text{V}_\text{N}\text{C}_\text{B}$, where the carbon atom of the defect relaxes out of plane. The out-of-plane configuration is stable compared to the planar configuration  and differs by 0.23 eV for monolayer hBN, while the energy difference of a defect in a monolayer compared to a defect embedded in the central hBN layer of a trilayer system amounts to  by 0.036 eV for $\text{V}_{\text{N}}\text{C}_{\text{B}}$ for example.
The analysis of phonon modes for $\text{V}_{\text{N}}\text{C}_{\text{B}}$ and $\text{C}_{\text{2}}\text{C}_{\text{N}}$ reveal that the bulk like modes have a dominant contribution for $\text{V}_{\text{N}}\text{C}_{\text{B}}$ while for $\text{C}_{\text{2}}\text{C}_{\text{N}}$ and $\text{C}_{\text{2}}\text{C}_{\text{B}}$ the localized stretch mode of the three carbon atoms has a dominant contribution to the PSB. This is in agreement with previous studies\cite{sajid_theoretical_2020, jara_first-principles_2021}.

Experimentally there are clearly two different spectral signatures of the inherent defects that are found in CVD hBN and LPE hBN samples. To quantify this difference and obtain more information on the respective electron to phonon coupling, we fit the data to the simulated spectra to extract the spectral width and DW factor. This is only performed for the narrow emission in LPE hBN and the emission from CVD hBN as these spectral features have been verified to stem from single photon emitters and hence single defects. 

When fitting the theory to our experimental data, we keep the relative spectral locations of the different transitions and the relative intensities of the different phonon modes fixed and allow the spectral width, the DW factor and the ZPL transition energy to vary. The latter is valid as DFT calculations are not accurate enough to determine the absolute transition energy but can be trusted to estimate the relative transition energies of the different modes involved. Further the absolute position of the ZPL can also be strongly affected by the defect's environment \cite{mendelson_strain-induced_2020}. The relative intensity of the ZPL compared to the total PL intensity is described by the DW factor and is also obtained by our simulations, where the values for the different defects are shown in table \ref{tab:DWFactor}. However, calculating
the DW factor for the exact experimental conditions is complicated as many factors such as stacking faults in the material can effect the contribution of light emitted into the ZPL \cite{lee_strong_2021}. A consistent discrepancy between simulated and measured DW factors, where the experimentally measured DW vastly exceeded the theoretical one has also been seen in other low-dimensional materials \cite{lee_strong_2021} and thus it seems reasonable to extract this value from the fit. 
\begin{table}
\centering
\begin{tabular}{|p{1.5cm}|p{3.5cm}|}
\hline
defect & DW$_\text{sim}$\\
\hline
\hline 
$\text{C}_\text{2}\text{C}_\text{B} $&0.47\\
\hline
$\text{V}_\text{N}\text{C}_\text{B}$ & 0.22\\
\hline
$\text{C}_\text{2}\text{C}_\text{N}$ (1) & 0.37\\
\hline
$\text{C}_\text{2}\text{C}_\text{N}$ (2)  & 0.26\\

\hline
\end{tabular}
\caption{Simulated DW factors for three different defects in hBN }
\label{tab:DWFactor}
\end{table}
The spectral width of the defects is affected by strain, temperature, spectral diffusion and electron-phonon coupling. Whereas a linear electron-phonon coupling leads to the PSB, a quadratic electron-phonon coupling leads to a homogeneous broadening contribution of the transitions. Spectral diffusion on the other hand will result in an inhomogeneous contribution to the width of the spectral lines \cite{jara_first-principles_2021}. To account for both factors, we initially fit the spectra to Voigt profiles. However, it is found that the inhomogeneous contribution at room temperature is insignificant compared to the homogeneous broadening as the Gaussian width yields a result consistent with zero. The fits in Fig. \ref{fig:expfit_theory_spectrum_lpe} and Fig. \ref{fig:expfit_theory_spectrum_cvd} are thus the results of a purely Lorentzian contribution and thus to 
\begin{equation}
A  \frac{(\gamma/2) ^2}{(x-x_\text{0,ZPL})^2+(\gamma/2)^2}+ A  \cdot f_\text{sc}\cdot \sum_{i=2}^{N} A_{\text{PSB},i}\frac{(\gamma/2) ^2}{(x-x_{0,\text{ZPL}}-x_{0,i})^2+(\gamma/2)^2},    
\end{equation}
where $x_{0,\text{ZPL}}$ represents the transition energy of the ZPL and $x_{0,i}$ are the relative spectral positions of the different modes as given by the simulations and fixed during the fit. $A_{\text{PSB},i}$ are the relative amplitudes of the different phonon modes. The fit parameters are the full width at half maximum $\gamma$, an overall amplitude $A$ and a scale factor $f_\text{sc}$  which accounts for the fact, that the DW factor may vary due to different experimental conditions. The extracted DW factor from the experiment is then related to the initially simulated DW factor $\text{DW}_{\text{sim}} = 1/(1+\sum_{i=2}^{N} A_{\text{PSB},i})$ by $\text{DW}_{\text{exp}}= 1/(1+f_{\text{sc}} \cdot (1-\text{DW}_{\text{sim}})/\text{DW}_{\text{sim}})$.

The fit results for the spectral width (FWHM) and experimentally obtained DW factor are summarized in Tab.\ref{tab:fitspectra}. Fig. \ref{fig:expfit_theory_spectrum_lpe} and Fig. \ref{fig:expfit_theory_spectrum_cvd} both show typical spectra and corresponding fits to the different simulated defects for emitters in the LPE and CVD sample, respectively. Below each spectrum the simulated transitions are also plotted.  From these fits we can immediately conclude that the $\text{V}_\text{N}\text{C}_\text{B}$ does not fit our data well and hence is unlikely to be the defect responsible for single photon emission of hBN in the visible range.

For the CVD sample, we find a good agreement with the  $\text{C}_\text{2}\text{C}_\text{B}$ defect. In fact, when looking closely at the data, this defect is also the only one with a significant contribution of phonon modes around 1.8 eV which results in the small third bump of the spectra and hence it seems to be a very likely candidate of single photon emission in these samples. During the preparation of this manuscript, we became aware of a complementary study \cite{fischer_combining_2023} at low-temperatures, which is also in good agreement with our results.

While we can also exclude the $\text{V}_\text{N}\text{C}_\text{B}$ defect for the LPE sample, both $\text{C}_\text{2}\text{C}_\text{B}$ and $\text{C}_\text{2}\text{C}_\text{N}$ could be possible candidates for this type of sample. However, it is clear that this indirect method of fitting our data to possible defects is more powerful in excluding candidates than pinpointing the exact defect and hence more complex direct correlative approaches may be necessary to gain certainty. Nevertheless we were able to narrow down the orign and also present a method that can be used to test for other possible suggested defects in the future. The results for our DW factors in both samples also show that they are very high even at room temperature. Such high DW factors are favorable compared to those of some other solid-state emitters at room temperature \cite{zhao_effect_2012} and may arise due to the weak linear electron-phonon coupling resulting from the 2D nature of the host material.

\begin{table}
\centering
\begin{tabular}{|p{1.5cm}||p{3.0cm}|p{3.0cm}||p{3.0 cm}|p{3.0cm}|}
\hline
 &\multicolumn{2}{|c||}{\text{CVD sample}}&\multicolumn{2}{c|}{\text{LPE sample (narrow spectrum)}}\\
&  $\gamma$ (FWHM) [eV] &  DW$_\text{exp}$ &  $\gamma$ (FWHM) [eV] &  DW$_\text{exp}$\\
\hline
\hline
C$_2$C$_\text{B}$ & $0.080\pm 0.016$&$0.74\pm 0.05$&$0.025\pm0.010$&$0.69\pm 0.02$\\
\hline
V$_\text{N}$C$_\text{B}$ & $0.068\pm 0.012$&$0.62\pm0.07$&$0.019\pm 0.005$&$0.51\pm0.05$\\
\hline
C$_2$C$_\text{N}$ (1) &$0.073\pm0.014$&$0.68\pm0.06$&$0.021\pm0.007$&$0.56\pm0.03$\\
\hline
C$_2$C$_\text{N}$ (2) & $0.090\pm0.034$&$0.78\pm0.13$&$0.022\pm0.008$&$0.55\pm0.03$\\
\hline
\end{tabular}
\caption{Average results of fitting the experimental spectra to the simulated defects. The errors are obtained by taking the standard deviation over several fits.}
\label{tab:fitspectra}
\end{table}

It is interesting to note that different sample preparations of the hBN host consistently lead to those distinct and reproducible properties of defects as shown here, which is also consistent with the observations of other research groups \cite{chejanovsky_structural_2016,abidi_selective_2019,comtet_wide-field_2019,glushkov_direct_2021, kim_integrated_2019,schell_quantum_2018,nguyen_nanoassembly_2018, zhong_carbon-related_2024, li_purification_2019, kretzschmar_quantitative_2024}. 

Stable emitters in CVD grown hBN usually exhibit larger spectral widths than those found in LPE samples. It has been proposed that emitters in CVD hBN may be more susceptible to spectral diffusion because of a smaller grain size \cite{vacha_inhomogeneous_1997} compared to LPE samples which may lead to more charge traps and it has been demonstrated that the overall linewidth can indeed be reduced by using a conductive substrate \cite{akbari_temperature-dependent_2021}. However, from our data we see that the broader linewidth is mainly due to a larger homogeneous broadening. This was further verified by fitting a Voigt profile to the CVD data with the homogeneous linewidth fixed to that of the LPE sample and letting the inhomogeneous linewidth float. This model fits less well than a Voigt profile with both contributions left floating which results in a Gaussian width consistent with zero and thus suggests a dominant contribution of the homogeneous part of the broadening. Such a broadening can be a result of a quadratic interaction between thermal acoustic phonons and the emitter \cite{akbari_temperature-dependent_2021}. Why this electron-phonon coupling should be stronger for CVD grown samples compared to LPE samples remains unclear. Another option is of course that these two types of emitters have a different atomic origin or that background fluorescence is a bigger problem in those samples. 

While it has been found that the linewidth of CVD-grown monolayers is generally larger than for multilayers due to shielding of the emitters from the environment \cite{tran_quantum_2016}, in a multilayer sample more layers do not necessarily mean even narrower linewidth. It has actually been demonstrated that annealing a multilayer CVD grown sample at higher temperatures and thus reducing the number of layers decreases the linewidth \cite{li_purification_2019}. In fact the authors of ref. 54 were able to show that broad emission typical for CVD grown samples could be converted to narrow emission characteristics reminiscent of the emission of emitters in LPE samples when annealing the samples at 750° in air for 3 hours. This annealing process showed to not only reduce the flake thickness but also the roughness, which the authors attributed to correlate with their best-quality single photon emitters. That the number of layers may not be the reason for the difference in linewidth is also underpinned by previous measurements at room and cryogenic temperatures where CVD grown and mechanically exfoliated samples of the same thickness were used \cite{akbari_temperature-dependent_2021}. 

\section{Conclusion}
In conclusion, we have investigated the emission characteristics of single photon emitters in hBN that inherently arise in samples produced by LPE and CVD. We have focused on nanofiber-integrated LPE nanocrystals and a layer-engineered trilayer CVD hBN sample. For the former, we have further demonstrated a simple method to interface hBN quantum emitters with optical nanofibers and hence optical fiber networks. 

That carbon can play an important role in the formation of some single photon emitters in hBN in the visible range has recently been demonstrated experimentally \cite{mendelson_identifying_2021,saha_comprehensive_2021, venturi_selective_2024}. However, the range of possible carbon-based emitters is still fairly large \cite{mendelson_identifying_2021,sajid_single-photon_2020}. On the route to being able to deterministically engineer these defects at desired locations, knowledge of the characteristics and the origin of the respective emitters is crucial. By fitting our experimental results of the emission characteristics to those of simulated results for different likely defects, we find that the $\text{C}_\text{2}\text{C}_\text{B}$ matches the results for the single photon emitters found in our CVD-grown samples best and that the $\text{V}_\text{N}\text{C}_\text{B}$ defect is unlikely to be the origin of the emitters that were found in CVD and LPE hBN.

The percentage of photon emission on the ZPL compared to emission into the PSB - the DW factor,  is one criterion to determine how useful a quantum emitter is for photonic quantum networks. Here we find DW factors around 70\% at room temperature for quantum emitters in both types of samples, which is very high for solid state quantum emitters\cite{zhao_suppression_2012,gorlitz_spectroscopic_2020,xue_experimental_2021} and underpins their suitability  for novel photonic quantum technologies.

Our experimental and theoretical work has narrowed down the number of candidates that may be responsible for single-photon emission in hBN and compares characteristics of emitters that are inherently present in samples that are produced very differently and that were chosen because of their suitablility for integration with nanophotonics. Thereby, we have obtained important results on the emitters' behaviour and properties which aids in understanding their nature and hence presents an important step towards integrated quantum networks and novel quantum technologies with hBN quantum emitters.

\section{Methods}\label{sec:methods}

\subsection{LPE sample}
The LPE sample used are LPE hBN nanoflakes that are suspended in a water:ethanol (45:55) mixture (concentration: $\tilde{}$ 5 mg/l ) \cite{coleman_two-dimensional_2011,griffin_spectroscopic_2018} suspension commercially sourced from Graphene Supermarket. 
To estimate the hBN flakes’ lateral size and thickness distributions and confirm their structure, atomic force microscopy (AFM, tapping mode, NT-MDT Ntegra Spectra), Raman spectroscopy (473 nm, NT-MDT Ntegra Spectra) and transmission electron microscopy (TEM, Philips CM200 at 80 kV electron acceleration voltage) in bright-field (BF) and dark-field (DF) modes and selected area electron diffraction (SAED) of hBN flakes drop cast onto planar SiO$_2$ (90 nm) covered Si wafers and lacey carbon TEM grids are used. The LPE sample is then interfaced with the light field by depositing it on an optical nanofiber or observed via our confocal microscope.

\subsection{Optical nanofiber set-up}
After production of the optical nanofiber by the heat-and-pull process \cite{warken_fiber_2008}, it is glued on an aluminium holder with the tapered region being free-standing.   The LPE sample is then interfaced with the light field by depositing it on the optical nanofiber via our "drop-touch" method. For spectral measurements or second-order intensity correlation measurements, the laser is sent through the optical fiber and backscattered fluorescence is separated from the excitation light via longpass filters and then either sent to the spectrometer (Shamrock SR-303i, Andor Technology) or a Hanbury-Brown-Twiss set-up with two single photon counting modules (SPCM-AQRH, Exelitas  Technologies).

\subsection{Layer-engineered CVD sample}
The individual layers are CVD grown and transferred onto a SiO$_2$/Si wafer using using a polyvinyl alcohol (PVA) support that can subsequently be dissolved in water. The layers are further cleaned by a post-transfer air-annealing step. This step also removes  all quantum emitters that have been created during the growth process. Only for the center layer a further Ar annealing step is applied to (re)activate and control the emitter density in the desired layer. These emitters are the same in nature as those that are initially inherently found in the CVD sample \cite{stewart_quantum_2021}. Hence, in the final sample, the central layer is sandwiched between two hBN layers that have been thoroughly cleaned by air annealing and function as a protection against bleaching of the quantum emitters.  Further details on sample preparation and characterisation of such samples can be found in Ref. 14.

\subsection{Confocal set-up}
 We study the samples using a homebuilt confocal microscope and a diode laser at 532 nm (Fig. \ref{fig:optical_setup_cvd}b). The excitation light is separated from the collected fluorescence by a dichroic mirror and two 550 nm long pass filters. The fluorescence is then either sent to a HBT set-up or to a spectrometer (Shamrock SR-303i, Andor Technology). We image the trilayer sample either in confocal or widefield mode.
For the widefield images, a single photon sensitive camera (iXon Ultra 897) is used. For obtaining confocal images, the sample is scanned by a piezo and the collected fluorescence is directed through a pinhole to the same single photon counting modules (SPCM-AQRH, Excelitas Technologies) which are also used for second order intensity correlation measurements.

\subsection{Second-order intensity correlation measurements}

For performing second-order intensity correlation measurements via the nanofiber interface, the backscattered fluorescence is sent via a symmetric beam splitter to two single photon counting modules (SPCM-AQRH, Exelitas  Technologies). A timetagger (Swabian Instruments) then records the single photon events. To eliminate excessive background scatter, which can be predominantly attributed to fluorescence and Raman scattering of the optical fiber, we use an additional shortpass filter for 650 nm for these measurements. For second order intensity correlation measurements via the confocal microscope, a dichroic mirror and a longpass filter at 550 nm is used to separate the excitation light from the fluorescence, which is directed through a beamsplitter to two single photon counting modules which are connected to a timetagger (Swabian Instruments). For both sample types the raw data is saved and then each photon event is correlated from sub-ns timescales to a few ms.

\subsection{Computational methods}
To simulate the PL spectra of different defects in hBN, calculations are performed for periodically replicated defects in a 2D hBN monolayer.  For calculation of total energy, electronic structure and ground state geometry we used version 5.3.3 of the Vienna Ab Initio Simulation Package (VASP) \cite{kresse_ab_1993,kresse_efficiency_1996}. We utilized the standard  projector augmented wave method (PAW)-projectors provided by the VASP package. Pristine single-layer hBN was first geometrically optimized using the conventional cell and a $27\times27\times1$ Monkhurst-Pack reciprocal space grid. A large vacuum region of 30 \AA $\,$width was used to separate a single layer of hBN from its periodic images and to ensure that interaction between periodic images is negligible. The optimized bond length of pristine hBN is 1.452 \AA. All the defects were then realized in a 9$\times$9$\times$1 supercell and allowed to fully relax using a plane wave cut-off of 700 eV for a maximum force of 0.001 eV \AA$^{-1}$. The k-point convergence was checked and finally a k-point mesh of $3\times3\times1$ was used for all calculations. The normal modes and dynamical matrices were calculated at Gamma point of the Brillouin zone. The total energies of the excited states were calculated within the $\Delta$SCF (self-consistent field) method that provides a reasonable estimate of ZPL energy and Stokes-shift for the optical excitation spectra for triplet \cite{sajid_defect_2018,abdulkadertawfik_first-principles_2017} and doublet \cite{sajid_theoretical_2020} manifolds. For further theory and computational details on calculations of the PL line shapes we refer to previous works\cite{sajid_vncb_2020}.
\section{Acknowledgments}

 S.M.S. acknowledges funding from the FETOpen program of the European Commission (no. 800942, ErBeStA ). S.M.S. and B.C.B acknowledge funding from the Austrian Research Promotion Agency (FFG) (no. 884447, PhoQus2D). S.M.S. is an  Elise Richter Fellow  of the Austrian Science Fund (FWF) (no. V934-N, Quantoom). K.S.T. is a Villum Investigator supported by VILLUM FONDEN (no. 37789). K.S.T. acknowledges funding from the European Research Council (ERC) under the European Union’s Horizon 2020 research and innovation program (no. 773122, LIMA). S.H. and V.B. acknowledge funding from EPSRC (EP/P005152/1) and the EU (Horizon 2020 Grant Agreement No. 785219). J.C.S. acknowledges support by the EPSRC Doctoral Training Centre in Graphene Technology (EP/L016087/1).

\section{Authors' contributions}

A. S. characterized LPE and CVD samples and performed data analysis on the results from all samples. H.S. interfaced the quantum emitters with the nanofiber and characterized the LPE hBN. B.C.B. undertook materials characterization of the LPE hBN. J.C.S, Y.F. and V.B. fabricated the layer-engineered hBN sample under the supervision of S.H. and pre-characterized it. S.A. and K.S.T. supplied the theoretical results for the different defects in hBN. S.W. and H.H. implemented the set-up for detection of single quantum emitters in layer-engineered hBN. A.S.P. and S.M.S. characterized the layer-engineered sample and analysed the data. B.C.B. and S.M.S conceived the idea for the experiments. S.M.S. supervised the project, analysed the data and wrote the manuscript with input from all the authors.

\section{Data availability}
The datasets used and/or analysed during the current study are available from the corresponding author on reasonable request.

\section{Competing interests}
The authors declare no competing interests.

\bibliography{quantumhbn,allRefNewNew,bibliography_hBNpaper2_new,bibliography_hBN_characteristics}

\begin{thebibliography}{10}
\urlstyle{rm}
\expandafter\ifx\csname url\endcsname\relax
  \def\url#1{\texttt{#1}}\fi
\expandafter\ifx\csname urlprefix\endcsname\relax\def\urlprefix{URL }\fi
\expandafter\ifx\csname doiprefix\endcsname\relax\def\doiprefix{DOI: }\fi
\providecommand{\bibinfo}[2]{#2}
\providecommand{\eprint}[2][]{\url{#2}}

\bibitem{tran_quantum_2016}
\bibinfo{author}{Tran, T.~T.}, \bibinfo{author}{Bray, K.}, \bibinfo{author}{Ford, M.~J.}, \bibinfo{author}{Toth, M.} \& \bibinfo{author}{Aharonovich, I.}
\newblock \bibinfo{journal}{\bibinfo{title}{Quantum emission from hexagonal boron nitride monolayers}}.
\newblock {\emph{\JournalTitle{Nature Nanotechnology}}} \textbf{\bibinfo{volume}{11}}, \bibinfo{pages}{37--41}, \doiprefix\url{10.1038/nnano.2015.242} (\bibinfo{year}{2016}).
\newblock \bibinfo{note}{Number: 1 Publisher: Nature Publishing Group}.

\bibitem{wang_site-specific_2024}
\bibinfo{author}{Wang, Y.}, \bibinfo{author}{Yu, Z.}, \bibinfo{author}{Smith, C.~S.} \& \bibinfo{author}{Caneva, S.}
\newblock \bibinfo{journal}{\bibinfo{title}{Site-{Specific} {Integration} of {Hexagonal} {Boron} {Nitride} {Quantum} {Emitters} on {2D} {DNA} {Origami} {Nanopores}}}.
\newblock {\emph{\JournalTitle{Nano Letters}}} \textbf{\bibinfo{volume}{24}}, \bibinfo{pages}{8510--8517}, \doiprefix\url{10.1021/acs.nanolett.4c00673} (\bibinfo{year}{2024}).
\newblock \bibinfo{note}{Publisher: American Chemical Society}.

\bibitem{caldwell_photonics_2019}
\bibinfo{author}{Caldwell, J.~D.} \emph{et~al.}
\newblock \bibinfo{journal}{\bibinfo{title}{Photonics with hexagonal boron nitride}}.
\newblock {\emph{\JournalTitle{Nature Reviews Materials}}} \textbf{\bibinfo{volume}{4}}, \bibinfo{pages}{552--567}, \doiprefix\url{10.1038/s41578-019-0124-1} (\bibinfo{year}{2019}).

\bibitem{gottscholl_sub-nanoscale_2021}
\bibinfo{author}{Gottscholl, A.} \emph{et~al.}
\newblock \bibinfo{journal}{\bibinfo{title}{Sub-nanoscale {Temperature}, {Magnetic} {Field} and {Pressure} sensing with {Spin} {Centers} in {2D} hexagonal {Boron} {Nitride}}}.
\newblock {\emph{\JournalTitle{arXiv:2102.10890 [cond-mat, physics:quant-ph]}}}  (\bibinfo{year}{2021}).
\newblock \bibinfo{note}{ArXiv: 2102.10890}.

\bibitem{vogl_room_2017}
\bibinfo{author}{Vogl, T.}, \bibinfo{author}{Lu, Y.} \& \bibinfo{author}{Lam, P.~K.}
\newblock \bibinfo{journal}{\bibinfo{title}{Room temperature single photon source using fiber-integrated hexagonal boron nitride}}.
\newblock {\emph{\JournalTitle{Journal of Physics D: Applied Physics}}} \textbf{\bibinfo{volume}{50}}, \bibinfo{pages}{295101}, \doiprefix\url{10.1088/1361-6463/aa7839} (\bibinfo{year}{2017}).

\bibitem{schell_non-linear_2016}
\bibinfo{author}{Schell, A.~W.}, \bibinfo{author}{Tran, T.~T.}, \bibinfo{author}{Takashima, H.}, \bibinfo{author}{Takeuchi, S.} \& \bibinfo{author}{Aharonovich, I.}
\newblock \bibinfo{journal}{\bibinfo{title}{Non-linear excitation of quantum emitters in hexagonal boron nitride multiplayers}}.
\newblock {\emph{\JournalTitle{APL Photonics}}} \textbf{\bibinfo{volume}{1}}, \bibinfo{pages}{091302}, \doiprefix\url{10.1063/1.4961684} (\bibinfo{year}{2016}).

\bibitem{kianinia_robust_2017}
\bibinfo{author}{Kianinia, M.} \emph{et~al.}
\newblock \bibinfo{journal}{\bibinfo{title}{Robust {Solid}-{State} {Quantum} {System} {Operating} at 800 {K}}}.
\newblock {\emph{\JournalTitle{ACS Photonics}}} \textbf{\bibinfo{volume}{4}}, \bibinfo{pages}{768--773}, \doiprefix\url{10.1021/acsphotonics.7b00086} (\bibinfo{year}{2017}).

\bibitem{liu_ultrastable_2020}
\bibinfo{author}{Liu, W.} \emph{et~al.}
\newblock \bibinfo{journal}{\bibinfo{title}{An ultrastable and robust single-photon emitter in hexagonal boron nitride}}.
\newblock {\emph{\JournalTitle{Physica E: Low-dimensional Systems and Nanostructures}}} \textbf{\bibinfo{volume}{124}}, \bibinfo{pages}{114251}, \doiprefix\url{10.1016/j.physe.2020.114251} (\bibinfo{year}{2020}).

\bibitem{fournier_position-controlled_2021}
\bibinfo{author}{Fournier, C.} \emph{et~al.}
\newblock \bibinfo{journal}{\bibinfo{title}{Position-controlled quantum emitters with reproducible emission wavelength in hexagonal boron nitride}}.
\newblock {\emph{\JournalTitle{Nature Communications}}} \textbf{\bibinfo{volume}{12}}, \bibinfo{pages}{3779}, \doiprefix\url{10.1038/s41467-021-24019-6} (\bibinfo{year}{2021}).
\newblock \bibinfo{note}{Number: 1 Publisher: Nature Publishing Group}.

\bibitem{nikolay_very_2019}
\bibinfo{author}{Nikolay, N.} \emph{et~al.}
\newblock \bibinfo{journal}{\bibinfo{title}{Very {Large} and {Reversible} {Stark}-{Shift} {Tuning} of {Single} {Emitters} in {Layered} {Hexagonal} {Boron} {Nitride}}}.
\newblock {\emph{\JournalTitle{Physical Review Applied}}} \textbf{\bibinfo{volume}{11}}, \bibinfo{pages}{041001}, \doiprefix\url{10.1103/PhysRevApplied.11.041001} (\bibinfo{year}{2019}).

\bibitem{hoese_mechanical_2020}
\bibinfo{author}{Hoese, M.} \emph{et~al.}
\newblock \bibinfo{journal}{\bibinfo{title}{Mechanical {Decoupling} of {Quantum} {Emitters} in {Hexagonal} {Boron} {Nitride} from {Low}-{Energy} {Phonon} {Modes}}}.
\newblock {\emph{\JournalTitle{arXiv:2004.10826 [cond-mat, physics:physics, physics:quant-ph]}}}  (\bibinfo{year}{2020}).
\newblock \bibinfo{note}{ArXiv: 2004.10826}.

\bibitem{dietrich_solid-state_2020}
\bibinfo{author}{Dietrich, A.}, \bibinfo{author}{Doherty, M.~W.}, \bibinfo{author}{Aharonovich, I.} \& \bibinfo{author}{Kubanek, A.}
\newblock \bibinfo{journal}{\bibinfo{title}{Solid-state single photon source with {Fourier} transform limited lines at room temperature}}.
\newblock {\emph{\JournalTitle{Physical Review B}}} \textbf{\bibinfo{volume}{101}}, \bibinfo{pages}{081401}, \doiprefix\url{10.1103/PhysRevB.101.081401} (\bibinfo{year}{2020}).
\newblock \bibinfo{note}{Publisher: American Physical Society}.

\bibitem{bourrellier_bright_2016}
\bibinfo{author}{Bourrellier, R.} \emph{et~al.}
\newblock \bibinfo{journal}{\bibinfo{title}{Bright {UV} {Single} {Photon} {Emission} at {Point} {Defects} in h-{BN}}}.
\newblock {\emph{\JournalTitle{Nano Letters}}} \textbf{\bibinfo{volume}{16}}, \bibinfo{pages}{4317--4321}, \doiprefix\url{10.1021/acs.nanolett.6b01368} (\bibinfo{year}{2016}).
\newblock \bibinfo{note}{Publisher: American Chemical Society}.

\bibitem{stewart_quantum_2021}
\bibinfo{author}{Stewart, J.~C.} \emph{et~al.}
\newblock \bibinfo{journal}{\bibinfo{title}{Quantum {Emitter} {Localization} in {Layer}-{Engineered} {Hexagonal} {Boron} {Nitride}}}.
\newblock {\emph{\JournalTitle{ACS Nano}}} \textbf{\bibinfo{volume}{15}}, \bibinfo{pages}{13591--13603}, \doiprefix\url{10.1021/acsnano.1c04467} (\bibinfo{year}{2021}).
\newblock \bibinfo{note}{Publisher: American Chemical Society}.

\bibitem{skoff_optical-nanofiber-based_2018}
\bibinfo{author}{Skoff, S.~M.}, \bibinfo{author}{Papencordt, D.}, \bibinfo{author}{Schauffert, H.}, \bibinfo{author}{Bayer, B.~C.} \& \bibinfo{author}{Rauschenbeutel, A.}
\newblock \bibinfo{journal}{\bibinfo{title}{Optical-nanofiber-based interface for single molecules}}.
\newblock {\emph{\JournalTitle{Physical Review A}}} \textbf{\bibinfo{volume}{97}}, \bibinfo{pages}{043839}, \doiprefix\url{10.1103/PhysRevA.97.043839} (\bibinfo{year}{2018}).

\bibitem{chejanovsky_structural_2016}
\bibinfo{author}{Chejanovsky, N.} \emph{et~al.}
\newblock \bibinfo{journal}{\bibinfo{title}{Structural {Attributes} and {Photodynamics} of {Visible} {Spectrum} {Quantum} {Emitters} in {Hexagonal} {Boron} {Nitride}}}.
\newblock {\emph{\JournalTitle{Nano Letters}}} \textbf{\bibinfo{volume}{16}}, \bibinfo{pages}{7037--7045}, \doiprefix\url{10.1021/acs.nanolett.6b03268} (\bibinfo{year}{2016}).

\bibitem{preus_assembly_2021}
\bibinfo{author}{Preuß, J.~A.} \emph{et~al.}
\newblock \bibinfo{journal}{\bibinfo{title}{Assembly of large {hBN} nanocrystal arrays for quantum light emission}}.
\newblock {\emph{\JournalTitle{2D Materials}}} \textbf{\bibinfo{volume}{8}}, \bibinfo{pages}{035005}, \doiprefix\url{10.1088/2053-1583/abeca2} (\bibinfo{year}{2021}).
\newblock \bibinfo{note}{Publisher: IOP Publishing}.

\bibitem{nguyen_nanoassembly_2018}
\bibinfo{author}{Nguyen, M.} \emph{et~al.}
\newblock \bibinfo{journal}{\bibinfo{title}{Nanoassembly of quantum emitters in hexagonal boron nitride and gold nanospheres}}.
\newblock {\emph{\JournalTitle{Nanoscale}}} \textbf{\bibinfo{volume}{10}}, \bibinfo{pages}{2267--2274}, \doiprefix\url{10.1039/C7NR08249E} (\bibinfo{year}{2018}).

\bibitem{kim_design_2018}
\bibinfo{author}{Kim, S.}, \bibinfo{author}{Toth, M.} \& \bibinfo{author}{Aharonovich, I.}
\newblock \bibinfo{journal}{\bibinfo{title}{Design of photonic microcavities in hexagonal boron nitride}}.
\newblock {\emph{\JournalTitle{Beilstein Journal of Nanotechnology}}} \textbf{\bibinfo{volume}{9}}, \bibinfo{pages}{102--108}, \doiprefix\url{10.3762/bjnano.9.12} (\bibinfo{year}{2018}).

\bibitem{schell_quantum_2018}
\bibinfo{author}{Schell, A.~W.}, \bibinfo{author}{Svedendahl, M.} \& \bibinfo{author}{Quidant, R.}
\newblock \bibinfo{journal}{\bibinfo{title}{Quantum {Emitters} in {Hexagonal} {Boron} {Nitride} {Have} {Spectrally} {Tunable} {Quantum} {Efficiency}}}.
\newblock {\emph{\JournalTitle{Advanced Materials}}} \textbf{\bibinfo{volume}{30}}, \bibinfo{pages}{1704237}, \doiprefix\url{https://doi.org/10.1002/adma.201704237} (\bibinfo{year}{2018}).
\newblock \bibinfo{note}{\_eprint: https://onlinelibrary.wiley.com/doi/pdf/10.1002/adma.201704237}.

\bibitem{mendelson_engineering_2019}
\bibinfo{author}{Mendelson, N.} \emph{et~al.}
\newblock \bibinfo{journal}{\bibinfo{title}{Engineering and {Tuning} of {Quantum} {Emitters} in {Few}-{Layer} {Hexagonal} {Boron} {Nitride}}}.
\newblock {\emph{\JournalTitle{ACS Nano}}} \textbf{\bibinfo{volume}{13}}, \bibinfo{pages}{3132--3140}, \doiprefix\url{10.1021/acsnano.8b08511} (\bibinfo{year}{2019}).
\newblock \bibinfo{note}{Publisher: American Chemical Society}.

\bibitem{caneva_nucleation_2015}
\bibinfo{author}{Caneva, S.} \emph{et~al.}
\newblock \bibinfo{journal}{\bibinfo{title}{Nucleation {Control} for {Large}, {Single} {Crystalline} {Domains} of {Monolayer} {Hexagonal} {Boron} {Nitride} via {Si}-{Doped} {Fe} {Catalysts}}}.
\newblock {\emph{\JournalTitle{Nano Letters}}} \textbf{\bibinfo{volume}{15}}, \bibinfo{pages}{1867--1875}, \doiprefix\url{10.1021/nl5046632} (\bibinfo{year}{2015}).
\newblock \bibinfo{note}{Publisher: American Chemical Society}.

\bibitem{bayer_introducing_2017}
\bibinfo{author}{Bayer, B.~C.} \emph{et~al.}
\newblock \bibinfo{journal}{\bibinfo{title}{Introducing {Overlapping} {Grain} {Boundaries} in {Chemical} {Vapor} {Deposited} {Hexagonal} {Boron} {Nitride} {Monolayer} {Films}}}.
\newblock {\emph{\JournalTitle{ACS Nano}}} \textbf{\bibinfo{volume}{11}}, \bibinfo{pages}{4521--4527}, \doiprefix\url{10.1021/acsnano.6b08315} (\bibinfo{year}{2017}).

\bibitem{kim_stacking_2013}
\bibinfo{author}{Kim, C.-J.} \emph{et~al.}
\newblock \bibinfo{journal}{\bibinfo{title}{Stacking {Order} {Dependent} {Second} {Harmonic} {Generation} and {Topological} {Defects} in h-{BN} {Bilayers}}}.
\newblock {\emph{\JournalTitle{Nano Letters}}} \textbf{\bibinfo{volume}{13}}, \bibinfo{pages}{5660--5665}, \doiprefix\url{10.1021/nl403328s} (\bibinfo{year}{2013}).
\newblock \bibinfo{note}{Publisher: American Chemical Society}.

\bibitem{li_prolonged_2023}
\bibinfo{author}{Li, S.~X.} \emph{et~al.}
\newblock \bibinfo{journal}{\bibinfo{title}{Prolonged photostability in hexagonal boron nitride quantum emitters}}.
\newblock {\emph{\JournalTitle{Communications Materials}}} \textbf{\bibinfo{volume}{4}}, \bibinfo{pages}{1--11}, \doiprefix\url{10.1038/s43246-023-00345-8} (\bibinfo{year}{2023}).
\newblock \bibinfo{note}{Number: 1 Publisher: Nature Publishing Group}.

\bibitem{zeng_single-photon_2024}
\bibinfo{author}{Zeng, L.} \emph{et~al.}
\newblock \bibinfo{journal}{\bibinfo{title}{Single-{Photon} {Emission} from {Point} {Defects} in {Hexagonal} {Boron} {Nitride} {Induced} by {Plasma} {Treatment}}}.
\newblock {\emph{\JournalTitle{ACS Applied Materials \& Interfaces}}} \textbf{\bibinfo{volume}{16}}, \bibinfo{pages}{24899--24907}, \doiprefix\url{10.1021/acsami.4c02601} (\bibinfo{year}{2024}).
\newblock \bibinfo{note}{Publisher: American Chemical Society}.

\bibitem{shafi_efficient_2020}
\bibinfo{author}{Shafi, K.~M.}, \bibinfo{author}{Nayak, K.~P.}, \bibinfo{author}{Miyanaga, A.} \& \bibinfo{author}{Hakuta, K.}
\newblock \bibinfo{journal}{\bibinfo{title}{Efficient fiber in-line single photon source based on colloidal single quantum dots on an optical nanofiber}}.
\newblock {\emph{\JournalTitle{Applied Physics B}}} \textbf{\bibinfo{volume}{126}}, \doiprefix\url{10.1007/s00340-020-7407-5} (\bibinfo{year}{2020}).

\bibitem{yalla_efficient_2012}
\bibinfo{author}{Yalla, R.}, \bibinfo{author}{Le~Kien, F.}, \bibinfo{author}{Morinaga, M.} \& \bibinfo{author}{Hakuta, K.}
\newblock \bibinfo{journal}{\bibinfo{title}{Efficient {Channeling} of {Fluorescence} {Photons} from {Single} {Quantum} {Dots} into {Guided} {Modes} of {Optical} {Nanofiber}}}.
\newblock {\emph{\JournalTitle{Phys. Rev. Lett.}}} \textbf{\bibinfo{volume}{109}}, \bibinfo{pages}{063602}, \doiprefix\url{10.1103/PhysRevLett.109.063602} (\bibinfo{year}{2012}).

\bibitem{schell_coupling_2017}
\bibinfo{author}{Schell, A.~W.}, \bibinfo{author}{Takashima, H.}, \bibinfo{author}{Tran, T.~T.}, \bibinfo{author}{Aharonovich, I.} \& \bibinfo{author}{Takeuchi, S.}
\newblock \bibinfo{journal}{\bibinfo{title}{Coupling {Quantum} {Emitters} in {2D} {Materials} with {Tapered} {Fibers}}}.
\newblock {\emph{\JournalTitle{ACS Photonics}}} \textbf{\bibinfo{volume}{4}}, \bibinfo{pages}{761--767}, \doiprefix\url{10.1021/acsphotonics.7b00025} (\bibinfo{year}{2017}).
\newblock \bibinfo{note}{Publisher: American Chemical Society}.

\bibitem{schell_highly_2015}
\bibinfo{author}{Schell, A.~W.} \emph{et~al.}
\newblock \bibinfo{journal}{\bibinfo{title}{Highly {Efficient} {Coupling} of {Nanolight} {Emitters} to a {Ultra}-{Wide} {Tunable} {Nanofibre} {Cavity}}}.
\newblock {\emph{\JournalTitle{Scientific Reports}}} \textbf{\bibinfo{volume}{5}}, \doiprefix\url{10.1038/srep09619} (\bibinfo{year}{2015}).

\bibitem{hutner_nanofiber-based_2020}
\bibinfo{author}{Hütner, J.} \emph{et~al.}
\newblock \bibinfo{journal}{\bibinfo{title}{Nanofiber-based high-{Q} microresonator for cryogenic applications}}.
\newblock {\emph{\JournalTitle{Optics Express}}} \textbf{\bibinfo{volume}{28}}, \bibinfo{pages}{3249}, \doiprefix\url{10.1364/OE.381286} (\bibinfo{year}{2020}).

\bibitem{takashima_fabrication_2019}
\bibinfo{author}{Takashima, H.} \emph{et~al.}
\newblock \bibinfo{journal}{\bibinfo{title}{Fabrication of a nanofiber {Bragg} cavity with high quality factor using a focused helium ion beam}}.
\newblock {\emph{\JournalTitle{Optics Express}}} \textbf{\bibinfo{volume}{27}}, \bibinfo{pages}{6792}, \doiprefix\url{10.1364/OE.27.006792} (\bibinfo{year}{2019}).

\bibitem{warken_fiber_2008}
\bibinfo{author}{Warken, F.}, \bibinfo{author}{Rauschenbeutel, A.} \& \bibinfo{author}{Bartholomaus, T.}
\newblock \bibinfo{journal}{\bibinfo{title}{Fiber {Pulling} {Profits} from {Precise} {Positioning}-{Precise} motion control improves manufacturing of fiber optical resonators}}.
\newblock {\emph{\JournalTitle{Photonics Spectra}}} \textbf{\bibinfo{volume}{42}}, \bibinfo{pages}{73} (\bibinfo{year}{2008}).

\bibitem{abidi_selective_2019}
\bibinfo{author}{Abidi, I.~H.} \emph{et~al.}
\newblock \bibinfo{journal}{\bibinfo{title}{Selective {Defect} {Formation} in {Hexagonal} {Boron} {Nitride}}}.
\newblock {\emph{\JournalTitle{Advanced Optical Materials}}} \textbf{\bibinfo{volume}{7}}, \bibinfo{pages}{1900397}, \doiprefix\url{10.1002/adom.201900397} (\bibinfo{year}{2019}).
\newblock \bibinfo{note}{Publisher: John Wiley \& Sons, Ltd}.

\bibitem{comtet_wide-field_2019}
\bibinfo{author}{Comtet, J.} \emph{et~al.}
\newblock \bibinfo{journal}{\bibinfo{title}{Wide-{Field} {Spectral} {Super}-{Resolution} {Mapping} of {Optically} {Active} {Defects} in {Hexagonal} {Boron} {Nitride}}}.
\newblock {\emph{\JournalTitle{Nano Letters}}} \textbf{\bibinfo{volume}{19}}, \bibinfo{pages}{2516--2523}, \doiprefix\url{10.1021/acs.nanolett.9b00178} (\bibinfo{year}{2019}).

\bibitem{glushkov_direct_2021}
\bibinfo{author}{Glushkov, E.} \emph{et~al.}
\newblock \bibinfo{journal}{\bibinfo{title}{Direct growth of hexagonal boron nitride on photonic chips for high-throughput characterization}}.
\newblock {\emph{\JournalTitle{arXiv:2103.15415 [cond-mat, physics:physics]}}}  (\bibinfo{year}{2021}).
\newblock \bibinfo{note}{ArXiv: 2103.15415}.

\bibitem{kim_integrated_2019}
\bibinfo{author}{Kim, S.} \emph{et~al.}
\newblock \bibinfo{journal}{\bibinfo{title}{Integrated on {Chip} {Platform} with {Quantum} {Emitters} in {Layered} {Materials}}}.
\newblock {\emph{\JournalTitle{Advanced Optical Materials}}} \textbf{\bibinfo{volume}{7}}, \bibinfo{pages}{1901132}, \doiprefix\url{10.1002/adom.201901132} (\bibinfo{year}{2019}).

\bibitem{kretzschmar_quantitative_2024}
\bibinfo{author}{Kretzschmar, T.} \emph{et~al.}
\newblock \bibinfo{journal}{\bibinfo{title}{Quantitative {Investigation} of {Quantum} {Emitter} {Yield} in {Drop}-{Casted} {Hexagonal} {Boron} {Nitride} {Nanoflakes}}}.
\newblock {\emph{\JournalTitle{ACS Applied Optical Materials}}} \textbf{\bibinfo{volume}{2}}, \bibinfo{pages}{1427--1435}, \doiprefix\url{10.1021/acsaom.4c00200} (\bibinfo{year}{2024}).
\newblock \bibinfo{note}{Publisher: American Chemical Society}.

\bibitem{sajid_defect_2018}
\bibinfo{author}{Sajid, A.}, \bibinfo{author}{Reimers, J.~R.} \& \bibinfo{author}{Ford, M.~J.}
\newblock \bibinfo{journal}{\bibinfo{title}{Defect states in hexagonal boron nitride: {Assignments} of observed properties and prediction of properties relevant to quantum computation}}.
\newblock {\emph{\JournalTitle{Physical Review B}}} \textbf{\bibinfo{volume}{97}}, \bibinfo{pages}{064101}, \doiprefix\url{10.1103/PhysRevB.97.064101} (\bibinfo{year}{2018}).
\newblock \bibinfo{note}{Publisher: American Physical Society}.

\bibitem{fischer_controlled_2021}
\bibinfo{author}{Fischer, M.} \emph{et~al.}
\newblock \bibinfo{journal}{\bibinfo{title}{Controlled generation of luminescent centers in hexagonal boron nitride by irradiation engineering}}.
\newblock {\emph{\JournalTitle{Science Advances}}} \textbf{\bibinfo{volume}{7}}, \bibinfo{pages}{eabe7138}, \doiprefix\url{10.1126/sciadv.abe7138} (\bibinfo{year}{2021}).
\newblock \bibinfo{note}{Publisher: American Association for the Advancement of Science Section: Research Article}.

\bibitem{mendelson_identifying_2021}
\bibinfo{author}{Mendelson, N.} \emph{et~al.}
\newblock \bibinfo{journal}{\bibinfo{title}{Identifying carbon as the source of visible single-photon emission from hexagonal boron nitride}}.
\newblock {\emph{\JournalTitle{Nature Materials}}} \textbf{\bibinfo{volume}{20}}, \bibinfo{pages}{321--328}, \doiprefix\url{10.1038/s41563-020-00850-y} (\bibinfo{year}{2021}).
\newblock \bibinfo{note}{Number: 3 Publisher: Nature Publishing Group}.

\bibitem{turiansky_boron_2021}
\bibinfo{author}{Turiansky, M.~E.} \& \bibinfo{author}{Van~de Walle, C.~G.}
\newblock \bibinfo{journal}{\bibinfo{title}{Boron dangling bonds in a monolayer of hexagonal boron nitride}}.
\newblock {\emph{\JournalTitle{Journal of Applied Physics}}} \textbf{\bibinfo{volume}{129}}, \bibinfo{pages}{064301}, \doiprefix\url{10.1063/5.0040780} (\bibinfo{year}{2021}).
\newblock \bibinfo{note}{Publisher: American Institute of Physics}.

\bibitem{gottscholl_initialization_2020}
\bibinfo{author}{Gottscholl, A.} \emph{et~al.}
\newblock \bibinfo{journal}{\bibinfo{title}{Initialization and read-out of intrinsic spin defects in a van der {Waals} crystal at room temperature}}.
\newblock {\emph{\JournalTitle{Nature Materials}}} \textbf{\bibinfo{volume}{19}}, \bibinfo{pages}{540--545}, \doiprefix\url{10.1038/s41563-020-0619-6} (\bibinfo{year}{2020}).
\newblock \bibinfo{note}{Number: 5 Publisher: Nature Publishing Group}.

\bibitem{toledo_electron_2018}
\bibinfo{author}{Toledo, J.~R.} \emph{et~al.}
\newblock \bibinfo{journal}{\bibinfo{title}{Electron paramagnetic resonance signature of point defects in neutron-irradiated hexagonal boron nitride}}.
\newblock {\emph{\JournalTitle{Physical Review B}}} \textbf{\bibinfo{volume}{98}}, \bibinfo{pages}{155203}, \doiprefix\url{10.1103/PhysRevB.98.155203} (\bibinfo{year}{2018}).
\newblock \bibinfo{note}{Publisher: American Physical Society}.

\bibitem{jara_first-principles_2021}
\bibinfo{author}{Jara, C.} \emph{et~al.}
\newblock \bibinfo{journal}{\bibinfo{title}{First-{Principles} {Identification} of {Single} {Photon} {Emitters} {Based} on {Carbon} {Clusters} in {Hexagonal} {Boron} {Nitride}}}.
\newblock {\emph{\JournalTitle{The Journal of Physical Chemistry A}}} \textbf{\bibinfo{volume}{125}}, \bibinfo{pages}{1325--1335}, \doiprefix\url{10.1021/acs.jpca.0c07339} (\bibinfo{year}{2021}).
\newblock \bibinfo{note}{Publisher: American Chemical Society}.

\bibitem{liu_quantum_2021}
\bibinfo{author}{Liu, H.} \emph{et~al.}
\newblock \bibinfo{title}{Quantum emitter formation in carbon-doped monolayer hexagonal boron nitride}.
\newblock \bibinfo{type}{Tech. Rep.} \bibinfo{number}{arXiv:2110.04780}, \bibinfo{institution}{arXiv} (\bibinfo{year}{2021}).
\newblock \doiprefix\url{10.48550/arXiv.2110.04780}.
\newblock \bibinfo{note}{ArXiv:2110.04780 [cond-mat, physics:physics] type: article}.

\bibitem{venturi_selective_2024}
\bibinfo{author}{Venturi, G.} \emph{et~al.}
\newblock \bibinfo{journal}{\bibinfo{title}{Selective {Generation} of {Luminescent} {Defects} in {Hexagonal} {Boron} {Nitride}}}.
\newblock {\emph{\JournalTitle{Laser \& Photonics Reviews}}} \textbf{\bibinfo{volume}{18}}, \bibinfo{pages}{2300973}, \doiprefix\url{10.1002/lpor.202300973} (\bibinfo{year}{2024}).
\newblock \bibinfo{note}{\_eprint: https://onlinelibrary.wiley.com/doi/pdf/10.1002/lpor.202300973}.

\bibitem{sajid_theoretical_2020}
\bibinfo{author}{Sajid, A.}, \bibinfo{author}{Reimers, J.~R.}, \bibinfo{author}{Kobayashi, R.} \& \bibinfo{author}{Ford, M.~J.}
\newblock \bibinfo{journal}{\bibinfo{title}{Theoretical spectroscopy of the \$\{{\textbackslash}mathrm\{{V}\}\}\_\{{\textbackslash}mathrm\{{N}\}\}\{{\textbackslash}mathrm\{{N}\}\}\_\{{\textbackslash}mathrm\{{B}\}\}\$ defect in hexagonal boron nitride}}.
\newblock {\emph{\JournalTitle{Physical Review B}}} \textbf{\bibinfo{volume}{102}}, \bibinfo{pages}{144104}, \doiprefix\url{10.1103/PhysRevB.102.144104} (\bibinfo{year}{2020}).
\newblock \bibinfo{note}{Publisher: American Physical Society}.

\bibitem{mendelson_strain-induced_2020}
\bibinfo{author}{Mendelson, N.}, \bibinfo{author}{Doherty, M.}, \bibinfo{author}{Toth, M.}, \bibinfo{author}{Aharonovich, I.} \& \bibinfo{author}{Tran, T.~T.}
\newblock \bibinfo{journal}{\bibinfo{title}{Strain-{Induced} {Modification} of the {Optical} {Characteristics} of {Quantum} {Emitters} in {Hexagonal} {Boron} {Nitride}}}.
\newblock {\emph{\JournalTitle{Advanced Materials}}} \textbf{\bibinfo{volume}{32}}, \bibinfo{pages}{1908316}, \doiprefix\url{10.1002/adma.201908316} (\bibinfo{year}{2020}).
\newblock \bibinfo{note}{\_eprint: https://onlinelibrary.wiley.com/doi/pdf/10.1002/adma.201908316}.

\bibitem{lee_strong_2021}
\bibinfo{author}{Lee, J.~H.} \emph{et~al.}
\newblock \bibinfo{journal}{\bibinfo{title}{Strong {Zero}-{Phonon} {Transition} from {Point} {Defect}-{Stacking} {Fault} {Complexes} in {Silicon} {Carbide} {Nanowires}}}.
\newblock {\emph{\JournalTitle{Nano Letters}}} \textbf{\bibinfo{volume}{21}}, \bibinfo{pages}{9187--9194}, \doiprefix\url{10.1021/acs.nanolett.1c03013} (\bibinfo{year}{2021}).
\newblock \bibinfo{note}{Publisher: American Chemical Society}.

\bibitem{fischer_combining_2023}
\bibinfo{author}{Fischer, M.} \emph{et~al.}
\newblock \bibinfo{journal}{\bibinfo{title}{Combining experiments on luminescent centres in hexagonal boron nitride with the polaron model and ab initio methods towards the identification of their microscopic origin}}.
\newblock {\emph{\JournalTitle{Nanoscale}}} \textbf{\bibinfo{volume}{15}}, \bibinfo{pages}{14215--14226}, \doiprefix\url{10.1039/D3NR01511D} (\bibinfo{year}{2023}).
\newblock \bibinfo{note}{Publisher: Royal Society of Chemistry}.

\bibitem{zhao_effect_2012}
\bibinfo{author}{Zhao, H.-Q.}, \bibinfo{author}{Fujiwara, M.} \& \bibinfo{author}{Takeuchi, S.}
\newblock \bibinfo{journal}{\bibinfo{title}{Effect of {Substrates} on the {Temperature} {Dependence} of {Fluorescence} {Spectra} of {Nitrogen} {Vacancy} {Centers} in {Diamond} {Nanocrystals}}}.
\newblock {\emph{\JournalTitle{Japanese Journal of Applied Physics}}} \textbf{\bibinfo{volume}{51}}, \bibinfo{pages}{090110}, \doiprefix\url{10.1143/JJAP.51.090110} (\bibinfo{year}{2012}).
\newblock \bibinfo{note}{Publisher: IOP Publishing}.

\bibitem{zhong_carbon-related_2024}
\bibinfo{author}{Zhong, D.} \emph{et~al.}
\newblock \bibinfo{journal}{\bibinfo{title}{Carbon-{Related} {Quantum} {Emitter} in {Hexagonal} {Boron} {Nitride} with {Homogeneous} {Energy} and 3-{Fold} {Polarization}}}.
\newblock {\emph{\JournalTitle{Nano Letters}}} \textbf{\bibinfo{volume}{24}}, \bibinfo{pages}{1106--1113}, \doiprefix\url{10.1021/acs.nanolett.3c03628} (\bibinfo{year}{2024}).
\newblock \bibinfo{note}{Publisher: American Chemical Society}.

\bibitem{li_purification_2019}
\bibinfo{author}{Li, C.} \emph{et~al.}
\newblock \bibinfo{journal}{\bibinfo{title}{Purification of single-photon emission from {hBN} using post-processing treatments}}.
\newblock {\emph{\JournalTitle{Nanophotonics}}} \textbf{\bibinfo{volume}{8}}, \bibinfo{pages}{2049--2055}, \doiprefix\url{10.1515/nanoph-2019-0099} (\bibinfo{year}{2019}).
\newblock \bibinfo{note}{Publisher: De Gruyter Section: Nanophotonics}.

\bibitem{vacha_inhomogeneous_1997}
\bibinfo{author}{Vacha, M.}, \bibinfo{author}{Liu, Y.}, \bibinfo{author}{Nakatsuka, H.} \& \bibinfo{author}{Tani, T.}
\newblock \bibinfo{journal}{\bibinfo{title}{Inhomogeneous and single molecule line broadening of terrylene in a series of crystalline n-alkanes}}.
\newblock {\emph{\JournalTitle{The Journal of Chemical Physics}}} \textbf{\bibinfo{volume}{106}}, \bibinfo{pages}{8324--8331}, \doiprefix\url{10.1063/1.473895} (\bibinfo{year}{1997}).

\bibitem{akbari_temperature-dependent_2021}
\bibinfo{author}{Akbari, H.}, \bibinfo{author}{Lin, W.-H.}, \bibinfo{author}{Vest, B.}, \bibinfo{author}{Jha, P.~K.} \& \bibinfo{author}{Atwater, H.~A.}
\newblock \bibinfo{journal}{\bibinfo{title}{Temperature-dependent {Spectral} {Emission} of {Hexagonal} {Boron} {Nitride} {Quantum} {Emitters} on {Conductive} and {Dielectric} {Substrates}}}.
\newblock {\emph{\JournalTitle{Physical Review Applied}}} \textbf{\bibinfo{volume}{15}}, \bibinfo{pages}{014036}, \doiprefix\url{10.1103/PhysRevApplied.15.014036} (\bibinfo{year}{2021}).
\newblock \bibinfo{note}{Publisher: American Physical Society}.

\bibitem{saha_comprehensive_2021}
\bibinfo{author}{Saha, S.} \emph{et~al.}
\newblock \bibinfo{journal}{\bibinfo{title}{Comprehensive characterization and analysis of hexagonal boron nitride on sapphire}}.
\newblock {\emph{\JournalTitle{AIP Advances}}} \textbf{\bibinfo{volume}{11}}, \bibinfo{pages}{055008}, \doiprefix\url{10.1063/5.0048578} (\bibinfo{year}{2021}).
\newblock \bibinfo{note}{Publisher: American Institute of Physics}.

\bibitem{sajid_single-photon_2020}
\bibinfo{author}{Sajid, A.}, \bibinfo{author}{Ford, M.~J.} \& \bibinfo{author}{Reimers, J.~R.}
\newblock \bibinfo{journal}{\bibinfo{title}{Single-photon emitters in hexagonal boron nitride: a review of progress}}.
\newblock {\emph{\JournalTitle{Reports on Progress in Physics}}} \textbf{\bibinfo{volume}{83}}, \bibinfo{pages}{044501}, \doiprefix\url{10.1088/1361-6633/ab6310} (\bibinfo{year}{2020}).
\newblock \bibinfo{note}{Publisher: IOP Publishing}.

\bibitem{zhao_suppression_2012}
\bibinfo{author}{Zhao, H.-Q.}, \bibinfo{author}{Fujiwara, M.} \& \bibinfo{author}{Takeuchi, S.}
\newblock \bibinfo{journal}{\bibinfo{title}{Suppression of fluorescence phonon sideband from nitrogen vacancy centers in diamond nanocrystals by substrate effect}}.
\newblock {\emph{\JournalTitle{Optics Express}}} \textbf{\bibinfo{volume}{20}}, \bibinfo{pages}{15628--15635}, \doiprefix\url{10.1364/OE.20.015628} (\bibinfo{year}{2012}).
\newblock \bibinfo{note}{Publisher: Optica Publishing Group}.

\bibitem{gorlitz_spectroscopic_2020}
\bibinfo{author}{Görlitz, J.} \emph{et~al.}
\newblock \bibinfo{journal}{\bibinfo{title}{Spectroscopic investigations of negatively charged tin-vacancy centres in diamond}}.
\newblock {\emph{\JournalTitle{New Journal of Physics}}} \textbf{\bibinfo{volume}{22}}, \bibinfo{pages}{013048}, \doiprefix\url{10.1088/1367-2630/ab6631} (\bibinfo{year}{2020}).
\newblock \bibinfo{note}{Publisher: IOP Publishing}.

\bibitem{xue_experimental_2021}
\bibinfo{author}{Xue, Y.} \emph{et~al.}
\newblock \bibinfo{journal}{\bibinfo{title}{Experimental {Optical} {Properties} of {Single}-{Photon} {Emitters} in {Aluminum} {Nitride} {Films}}}.
\newblock {\emph{\JournalTitle{The Journal of Physical Chemistry C}}} \textbf{\bibinfo{volume}{125}}, \bibinfo{pages}{11043--11047}, \doiprefix\url{10.1021/acs.jpcc.1c01376} (\bibinfo{year}{2021}).
\newblock \bibinfo{note}{Publisher: American Chemical Society}.

\bibitem{coleman_two-dimensional_2011}
\bibinfo{author}{Coleman, J.~N.} \emph{et~al.}
\newblock \bibinfo{journal}{\bibinfo{title}{Two-{Dimensional} {Nanosheets} {Produced} by {Liquid} {Exfoliation} of {Layered} {Materials}}}.
\newblock {\emph{\JournalTitle{Science}}} \textbf{\bibinfo{volume}{331}}, \bibinfo{pages}{568--571}, \doiprefix\url{10.1126/science.1194975} (\bibinfo{year}{2011}).
\newblock \bibinfo{note}{Publisher: American Association for the Advancement of Science}.

\bibitem{griffin_spectroscopic_2018}
\bibinfo{author}{Griffin, A.} \emph{et~al.}
\newblock \bibinfo{journal}{\bibinfo{title}{Spectroscopic {Size} and {Thickness} {Metrics} for {Liquid}-{Exfoliated} h-{BN}}}.
\newblock {\emph{\JournalTitle{Chemistry of Materials}}} \textbf{\bibinfo{volume}{30}}, \bibinfo{pages}{1998--2005}, \doiprefix\url{10.1021/acs.chemmater.7b05188} (\bibinfo{year}{2018}).
\newblock \bibinfo{note}{Publisher: American Chemical Society}.

\bibitem{kresse_ab_1993}
\bibinfo{author}{Kresse, G.} \& \bibinfo{author}{Hafner, J.}
\newblock \bibinfo{journal}{\bibinfo{title}{Ab initio molecular dynamics for liquid metals}}.
\newblock {\emph{\JournalTitle{Physical Review B}}} \textbf{\bibinfo{volume}{47}}, \bibinfo{pages}{558--561}, \doiprefix\url{10.1103/PhysRevB.47.558} (\bibinfo{year}{1993}).
\newblock \bibinfo{note}{Publisher: American Physical Society}.

\bibitem{kresse_efficiency_1996}
\bibinfo{author}{Kresse, G.} \& \bibinfo{author}{Furthmüller, J.}
\newblock \bibinfo{journal}{\bibinfo{title}{Efficiency of ab-initio total energy calculations for metals and semiconductors using a plane-wave basis set}}.
\newblock {\emph{\JournalTitle{Computational Materials Science}}} \textbf{\bibinfo{volume}{6}}, \bibinfo{pages}{15--50}, \doiprefix\url{10.1016/0927-0256(96)00008-0} (\bibinfo{year}{1996}).

\bibitem{abdulkadertawfik_first-principles_2017}
\bibinfo{author}{Abdulkader Tawfik, S.} \emph{et~al.}
\newblock \bibinfo{journal}{\bibinfo{title}{First-principles investigation of quantum emission from {hBN} defects}}.
\newblock {\emph{\JournalTitle{Nanoscale}}} \textbf{\bibinfo{volume}{9}}, \bibinfo{pages}{13575--13582}, \doiprefix\url{10.1039/C7NR04270A} (\bibinfo{year}{2017}).
\newblock \bibinfo{note}{Publisher: Royal Society of Chemistry}.

\bibitem{sajid_vncb_2020}
\bibinfo{author}{Sajid, A.} \& \bibinfo{author}{Thygesen, K.~S.}
\newblock \bibinfo{journal}{\bibinfo{title}{{VNCB} defect as source of single photon emission from hexagonal boron nitride}}.
\newblock {\emph{\JournalTitle{2D Materials}}} \textbf{\bibinfo{volume}{7}}, \bibinfo{pages}{031007}, \doiprefix\url{10.1088/2053-1583/ab8f61} (\bibinfo{year}{2020}).
\newblock \bibinfo{note}{Publisher: IOP Publishing}.

\end{thebibliography}

\end{document}